\documentclass[12pt]{article}
\usepackage{amsfonts}
\usepackage{amssymb}
\usepackage[utf8]{inputenc}
\usepackage{graphicx,color}
\usepackage[vlined,figure]{algorithm2e}

\usepackage{geometry}
\newcommand{\beqn}{\begin{eqnarray}}
\newcommand{\eeqn}{\end{eqnarray}}
\newcommand{\beq}{\begin{equation}}
\newcommand{\eeq}{\end{equation}}

\newcommand{\rf}{{\mathrm{RF}}}
\newcommand{\pS}{{\it pSeven~}}
\newcommand{\mns}[1]{\scalebox{0.5}[1.0]{\(-\)}#1}

\graphicspath{{./figs/}}
\title{
Towards an Automated Optimization\\
of Laminated Composite Structures:\\
Hierarchical Zoning Approach with\\
Exact Blending Rules}

\author{D.T.~Shvarts, ~ F.V.~Gubarev}

\date{
\small{DATADVANCE LLC},\\
\footnotesize{Russia 117246, Nauchny pr.~17, 15~fl., Moscow}
}
\begin{document}
\maketitle
\thispagestyle{empty}
\begin{abstract}
\noindent
We present an automated methodology to optimize laminated composite structures.
Our approach, inspired by bi-level optimization scheme, avoids its prime deficiencies
and is based on developed computationally inexpensive stacking sequences reconstruction algorithm,
which identically satisfies conventional set of blending rules.
We combine it with proposed tight approximation to feasible domain of composite integral parameters
and hierarchical zoning procedure to get highly efficient optimization methodology,
which we test in two example applications.
In both cases it is shown that blending rules compliant optimal solution remains within 10\% gap from
one with no rules applied.
\end{abstract}
\newpage
\section{Introduction and Overview}
\label{section::intro}
Design of an optimal composite structure is the problem to find the best
(usually, minimal weight) distribution of locally variable thicknesses
and fiber orientations (stacking sequences), which respect structural mechanical
constraints and manufacturing (blending) rules as well as some other restrictions,
e.g. related to damage-tolerances.
It is important to distinguish usual constraints,
which quantitatively characterize the degree of their violation by each particular design,
and various rules, which are similar to binary markers of design feasibility and for which
the notion of violation degree makes no sense.
Usual constraints are to be accounted for in conventional optimization-like treatment.
On the other hand, blending rules, which we discuss in length in Section~\ref{section::rules},
require the problem itself to be formulated such that possibility
of their violation never arises.

Complexity of optimal design process might be illustrated already at the level
of usual constraints. Indeed, evaluation of mechanical properties of composite structure
is often computationally intensive and requires sophisticated modeling software. 
This means that one is required to solve large-scale, expensive to evaluate,
non-linear, constrained mixed-integer optimization problem, which is,
in fact, not possible at all.
The remedy is to get some insights into the problem structure, which would allow to apply
practically relevant (albeit not rigorous) approximation schemes.
Crucial points to be exploited are:
\begin{enumerate}

\item Presence of small parameter.\\
In virtually all industrial applications the number of plies $n_i$, laid up
at each admissible angle $\theta_\alpha$, $\alpha=1,...,N_\theta$, is large,
which is the direct consequence of small single ply thickness.
Note that in this paper we exclusively consider the most common case of
symmetric balanced laminates with $N_\theta = 4$ possible ply orientations $\theta_\alpha \in \{ 0, \pm 45, 90 \}$,
although our approach applies equally in more general settings.
Therefore the problem actually possesses small
parameter $1/N \ll 1 $, which allows to treat relative number of plies $\rho_\alpha = n_\alpha/N$
(percentages) as continuous non-negative variables.
How large $N$ should be in order to justify the relation $1/N \ll 1$?
Experience revealed that already moderate values $N \sim 10$ are, in fact,
very close to that limit and therefore the above reasoning applies to virtually
all practically relevant cases.

\item Structural problem decomposition.\\
It is crucial that mechanical constraints depend upon specific
stacking sequences only indirectly, via the (half-)total number of plies $N$, percentages $\rho$
and the set of integrated composite material characteristics $\xi$, known as lamination parameters.
Note that there are exponentially large number of specific stacking sequences,
which correspond to the same $(N, \rho, \xi )$-values. Therefore, assuming that
performance criteria could be reformulated in terms of these parameters as well
(which is not always the case, see below) one could structurally decompose the original problem
into two levels (bi-level scheme, detailed, for instance, in
Refs.~\cite{toropov1,toropov2,Dimitri_Bettebghor,self}; for a review
of recent developments see, e.g., Refs.~\cite{phd1,phd2}): \\
{\it i)} Global macroscopic level aimed to find optimal $(N, \rho, \xi )$ values.
Here one works exclusively with macroscopic characteristics with no regard to actual stacking sequences.
The puspose  is to satisfy structural mechanical constraints and to provide best attainable performances.
It is important that the resulting problem is usual continuous constrained optimization
task admitting efficient solution strategies and all the computationally
intensive activities are concentrated only at this level.\\
{\it ii)} Local microscopic level intended to establish particular stacking sequences, which provide
$(N,\rho,\xi)$ parameters values closest to ones, obtained at global level.
It is of crucial importance that there are exponentially large number of corresponding
solutions~\cite{Dimitri_Bettebghor}. Moreover, despite of imposed manufacturing
and integrity constraints (blending and composition rules), the number of realizing stacking sequences
remains exponentially large and provides a kind theoretical justification to the methodology
presented in this paper.
\end{enumerate}

\noindent
As far as the upper (global) level problem is concerned, parameters to be considered
in case of balanced symmetric laminates are:
\begin{enumerate}
\item Number of plies $n_\alpha$, $\alpha = 0, \pm 45, 90$, laid up at different orientations
(or equivalently the corresponding percentages $\rho_\alpha = n_\alpha / N$) and the half-total number of plies $N$.
\item Integral characteristics of laminate composition,
      the so called lamination parameters $\xi_i$, $i=1,2,3$, defined as:
\beq
\label{xi-def}
\begin{array}{c}
\vec{\xi} ~=~ 
3 \, h^{-3} \int_0^h\, dz \,\, \vec{\zeta}(z) \,\, z^2 ~=~
3 \, h^{-3} \, \sum^N_{k=1} \vec{\zeta}(z_k) \ z^2_k \ \Delta z \,\, \in \, [-1;1]^3\\
~\\
\vec{\zeta}(z_k) ~=~ [\,\, \cos(2\theta_k), \,\, \cos(4\theta_k), \,\,\sin(2\theta_k)\,\,]^T\,.
\end{array}
\eeq
Here $\Delta z$ denotes one ply thickness with $h = N \Delta z$
being the panel half-width, $\theta_k$ is orientation of $k$-th ply, $z_k = k \, \Delta z$ is
the distance of $k$-th lamina to mid-plane and we used vector notations instead of explicit indices.
Note that for considered orientations $\vec{\xi}$ has only three components.
\end{enumerate}
It is worth noting that nowadays there are several efficient methodologies available to deal with
global level problem formulation. It is extremely important to establish exact feasible region of
macroscopic parameters, which warrants the existence of realizing stacking sequences.
It is known~\cite{domain1,domain2,domain3}
that for given percentages $\rho$ feasible region of $\xi$-parameters is a convex polytope
defined by explicit set of linear constraints. This fact was utilized in Ref.~\cite{self} to argue that
the usage of collaborative strategies supplemented with standard optimization methods might greatly
facilitate the solution process. However, collaborative schemes are generically expected to be
less efficient than properly constructed direct approaches. We address this issue in Section~\ref{section::domain}
and develop tight convex approximation to feasible region of combined $(\rho,\xi)$ parameters,
which allows us to apply efficient direct optimization strategies.

It is hardly possible to underestimate the advantages of the above partitioning scheme, which
provides a clear separation of scales in the problem.
Indeed, performance of designed macroscopic structure mostly depends upon its macroscopic
characteristics and not on its microscopic composition. 
Moreover, exponentially large number of stacking sequences corresponding to the same values
of $(N,\rho,\xi)$ parameters is reminiscent to the notion of entropy in statistical physics.
Given that we are confident that bi-level scheme is fruitful concept and is to be preserved
in future developments of composites optimization methods.
Unfortunately, in its original formulation presented above it has rather limited applicability.
Point is that major part of modeling tools used to predict performances of composite structures
are in-house developed and, being provided with complete set of specific stacking sequences,
evaluate at once both large and small scale characteristics.
Relevant microscopic observables of interest might include,
for instance, interlaminar shear stresses, strain check on every ply, etc.
Therefore, application of bi-level methodology is often impossible because real-life applications
might not technically admit explicit scales separation.

In this paper we solve the above issue via the development of fast algorithm to establish
representative set of stacking sequences, corresponding to given values of macroscopic parameters.
Moreover, reconstructed stacking sequences identically satisfy conventional set of blending
and composition rules, which we present and discuss in length in Section~\ref{section::SLB}.
Therefore, the algorithm essentially solves the problem posed at second (local) level of bi-level scheme,
but it is computationally inexpensive and thus could be embedded into the upper level solution scheme.
Although the methodology then ceases to be bi-level, it yet retains prime advantages of
bi-level approach: optimization is performed in terms of macroscopic parameters only,
representative stacking sequences are reconstructed ``on the fly'' essentially utilizing 
the fact that there are exponentially large number of appropriate solutions.

Important ingredient of our methodology is the automated hierarchical zoning.
Indeed, best performance could ever be achieved only if
macroscopic parameters are allowed to be spatially inhomogeneous.
In many cases design of large composite structures is based on its subdivision
into local panel (or zone) design problems, in which the material composition of each panel is
determined under assumed fixed loads.
Since different zones are structurally interrelated the process is to be iterated:
refined material composition in particular zone affects the loads of other panels
and hence requires to solve panel-based problem anew.
Even within such iterative settings it remains unclear how to properly select the number and locations
of different zones. Inadequate choice might result in vast number of design variables,
for which the problem might become intractable.
We propose adaptive multi-scale methodology (Section~\ref{section::zoning}),
which was proven to be extremely useful in various contexts.
It allows not only to automatically select appropriate number of zones, but also
significantly facilitates the solution process.
Namely, one starts with only a few (perhaps, a single) zones, for which the solution might be
found easily although it is expected to be far from being really optimal.
Then the obtained zones are subdivided further and problem is solved anew with smaller 
spatial resolution using previous result as the initial guess.
Process stops when either smallest allowed resolution is
reached or sequentially achieved gain in performances becomes negligible.

Despite of its ideological simplicity the hierarchical zoning outlined above is technically
involved. For instance, in real-life applications the number of zones to be considered simultaneously
might be large (a few decades, say). Corresponding total number of macroscopic parameters
is then of order $O(100)$, making optimization problem almost intractable.
To circumvent this we propose to use locality of zones interaction in physically motivated problems.
Indeed, the prime reason of rapidly rising dimensionality is structural interconnection
of different zones.
If there would be no loads redistribution, problem factorizes into the set of independent optimization tasks
for each zone.
On the other hand, change of the properties of particular zone affects only its neighbors,
for sufficiently distant panels loads remain practically the same.
This suggests the possibility to develop a specific optimization method, which exploits the locality
of zone interactions and remains efficient even when number of zones is large.
In Section~\ref{section::method} corresponding implementation is discussed in details.

Methodology and algorithms outlined above were applied to well-known horseshoe and wing box composite
structure optimization problems, for which a number of results obtained with other methods are available.
It is convenient that both test cases might be solved without blending rules imposition (they both
admit conventional bi-level treatment) thus providing a reference value, which is a lower bound
on attainable mass. 
It turns out that with exact accounting for composition rules optimal mass remains within $\sim 10\%$
gap from the respective lower bound.
This result is encouraging because {\it a priori} one could expect that 
composition rules might induce severe non-local interactions between different zones 
and hence highly degrade attainable performances.
However, this turns out to be not the case.
To the contrary, we show that proper accounting for blending rules opens the possibility
to design technologically compliant structures with almost theoretically attainable performances.

\section{Blending Rules}
\label{section::rules}

Manufacturing and integrity constraints, commonly known as blending or composition rules,
play a central role in composite structures optimization. It is crucial that they are not
the constraints {\it per se} because it is not possible to quantitatively characterize
the degree of their violation. Instead they should be considered as a set of binary
classifiers, which mark infeasibility of a particular design.
Moreover, if the design appears to be infeasible with respect to blending rules it is hardly
possible to point out how its stacking composition could be ameliorated.
Therefore, the term ``constraints'' should not be considered literally, blending rules
are indeed ``the rules'' and are to be satisfied identically.

Stacking composition rules don't have functional form and are usually
tabulated in the form of requirements imposed on admissible stacking sequences.
For ply orientations $\{ 0, \pm 45, 90 \}$ considered in this paper
conventional set of blending rules could be summarized as follows~\cite{blending1,blending2}:
\begin{itemize}
\item[1.] Laminate is to be symmetric and balanced;
\item[2.] Minimal allowed relative number (percentage) of plies of each orientation is 8\%;
\item[{\bf 3}.] At most 3 plies of same orientation could be grouped sequentially together
  ({\it contiguity});
\item[{\bf 4}.] Change of orientation between neighboring plies must not exceed $\pm 45$
  ({\it disorientation});
\item[5.] Orientation angle of outer-most plies is to be $\pm 45$;
\item[6.] Outer-most plies must be continuous through all the structure (no drop-off);
\item[{\bf 7}.] Continuity of ply orientation between zones must be respected
  ({\it continuity});
\item[{\bf 8}.] For every 3 dropped plies there must be at least one continuous
  ply covering the dropped ones ({\it dropping}).
\end{itemize}
Note that major part of the imposed rules might easily be accomplished in the considered approach.
Indeed, optimization methodology to be detailed in Section~\ref{section::method} operates
with macroscopic design variables, which are the (half-)number of plies $N$, percentages $\rho$ and laminaion parameters $\xi$.
Balance and symmetry requirements become trivial once we identify $n_{45} = n_{-45} = n_{\pm 45}$
and consider only the half of laminate width. Rules 5 and 6 might be identically accomplished
by reservation of required continuous plies at very beginning.
Therefore, the only non-trivial rules to be considered are the rules 3,4 and 7,8 marked in bold in the above list.
Grouping of them in two pairs is not occasional: contiguity and disorientation rules
refer to each particular zone with no regards to its possible neighbors and are to be taken into account always.
As for continuity and dropping rules, they operate on the boundaries of neighboring
zones and become void when only a single zone is considered.

It is essential that only continuity and dropping rules deserve specific attention. This is because it is
well-known (see, e.g., Refs.~\cite{MILP1,MILP2} and references therein)
how contiguity and disorientation requirements might be accounted for in single zone settings
via linear integer programming formulation of stacking sequence reconstruction problem.
Since we are going to exploit similar ideas in multi-zoning context it makes sense to briefly
review the most essential points.

\begin{figure}[t]
\centerline{\includegraphics[width=0.5\textwidth]{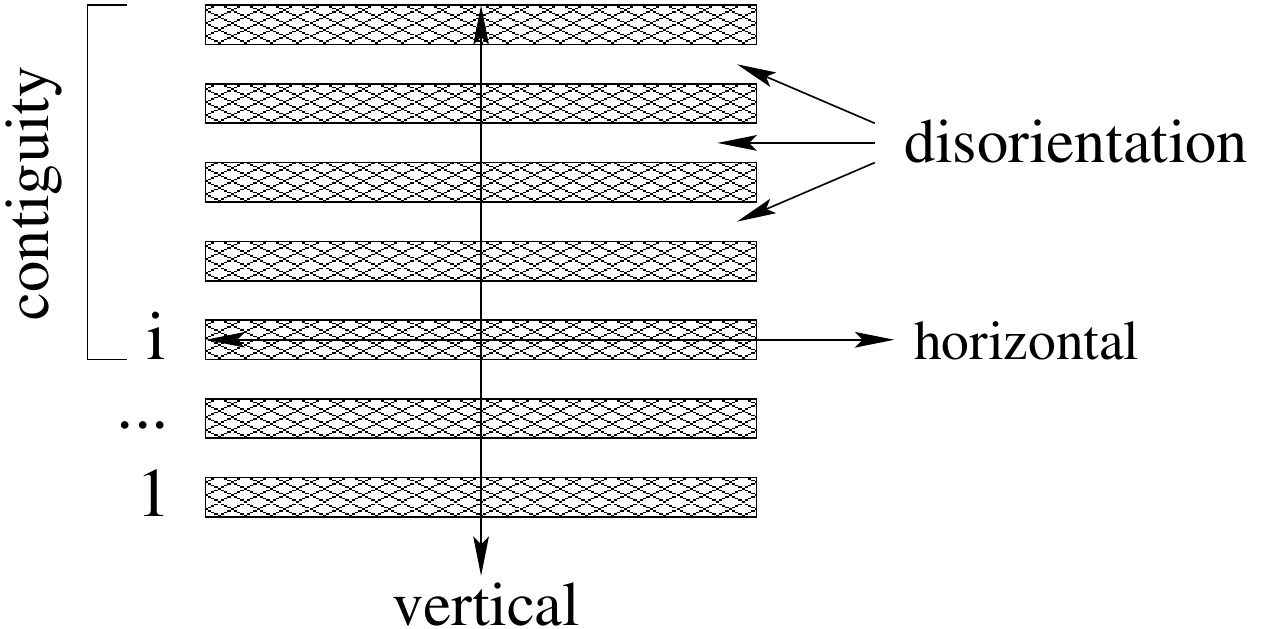}}
\caption{Single zone stacking sequence reconstruction: illustration of ``horizontal''/``vertical''
constraints and contiguity/disorientation rules implementation.}
\label{fig::horizontal-vertical}
\end{figure}

For single zone one is given the (half-)total number of plies $N$ and the corresponding percentages $\rho_\alpha$
or equivalently the number of plies $n_\alpha$ laid up at particular angle, which constitutes
the so called zone's signature vector $\vec{n}$.
For each ply with index $i$ from mid-plane one introduces a set of binary variables $s_{i,\alpha} = \{0,1\}$
describing its orientation, subject to obvious ``horizontal'' constraints
\beq
\label{horizontal}
\sum_\alpha \, s_{i,\alpha} ~=~ 1 \qquad \forall i \in [1,N]
\eeq
and  a set of ``vertical'' relations
\beq
\label{vertical}
\sum_i \, s_{i,\alpha} ~=~ n_\alpha \,,
\eeq
which enforce stacking sequence to have given signature
(distinction between ``horizontal'' and ``vertical'' constraints and qualitative explanation of
corresponding mnemonics is illustrated on Fig.~\ref{fig::horizontal-vertical}).
In term of binary variables  contiguity and disorientation rules become an additional set of linear
constraints operating either in bulk of laminate (contiguity) or on the interfaces (disorientation) of neighboring
plies~\cite{MILP1,MILP2}.
To complete the setup one has to specify the objective function, which is worth to take linear in $s_{i,\alpha}$
in order to maintain problem complexity.
Various performance measures could be used (for instance, uniformity of plies distribution through the stack),
however, in our approach the natural goal is to fit externally given lamination parameters values $\xi^*$.
Therefore, the objective function to consider is
\beq
\label{objective}
f ~=~ \sum_i \gamma_i\,,
\eeq
where additional variables $\gamma_i \ge 0$  and corresponding constraints
\beq
\label{gamma-constraints}
|\xi_i(s) ~-~ \xi_i^*| ~ \le ~ \gamma_i
\eeq
are used to represent $L_1$ norm of $\xi$-residual as a linear function.

A few comments are now in order. In the above formulation total number of plies and zone signature are preserved
by construction.
On the other hand, for any finite $N$ it is not possible to reproduce
externally given continuous values $\xi^*$ exactly, the discrepancy is always of order $1/N$.
However, as far as stiffnesses are concerned, there is no much difference between percentages $\rho$
and lamination parameters $\xi$, both parameterize rather symmetrically in-plane and out-of-plane stiffness matrices.
Then the apparent distinction between $\rho$ and $\xi$ expressed
in (\ref{vertical}-\ref{gamma-constraints})
appears unnatural, it is always worth to preserve problem symmetries and  to consider $\rho$, $\xi$
variables on equal footing. To achieve this we alter (\ref{vertical}), (\ref{objective}) to read
\beq
\label{vertical2}
|\sum_i \, s_{i,\alpha} ~-~ n_\alpha| ~ \le ~ N \, \kappa_\alpha\,,
\eeq
\beq
\label{objective2}
f ~=~ \sum_i \, \gamma_i ~+~ \sum_\alpha \kappa_\alpha\,,
\eeq
where the additional variables $\kappa_\alpha$ are non-negative and
the factor $N$ is required to have proper scales of different contributions.
Therefore, single zone stacking sequence reconstruction might be performed via the solution
of linear integer optimization problems of two types:
\begin{enumerate}
\item The one defined in (\ref{horizontal}-\ref{gamma-constraints}), which treats $\rho$, $\xi$ variables differently;
\label{linear-problem1}
\item The solution of (\ref{horizontal},\ref{vertical2},\ref{objective2}), where
  percentages and lamination parameters are considered symmetrically.
\label{linear-problem2}
\end{enumerate}
Second formulation is advantageous not only because of symmetry, but also because it is much more flexible with respect
to problem feasibility.
Both formulations are used below and are to be supplemented with identical set of constraints related
to contiguity and disorientation rules, which we do not present because of their simplicity.

\begin{figure}[t]
\centerline{\includegraphics[width=0.9\textwidth]{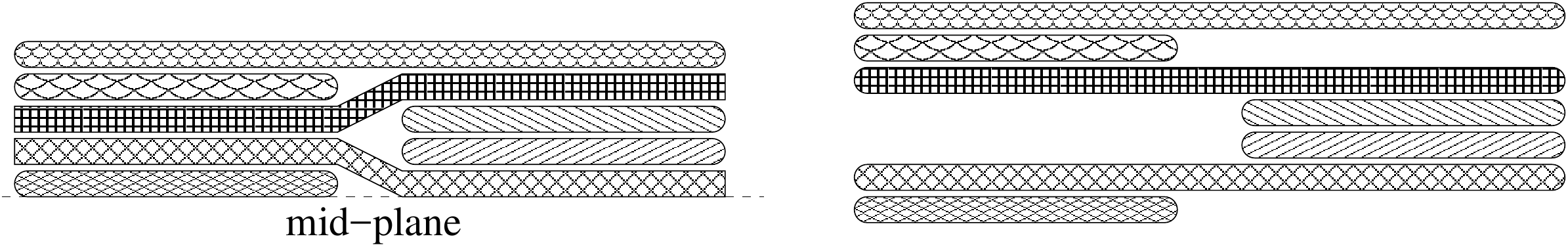}}
\caption{Patch set representation for composite structure with two zones.}
\label{fig::patch}
\end{figure}

Determination of stacking sequences for multi-zoned structure differs from above
single zone treatment in several aspects.
First, there are plies common to different zones, for which corresponding $s_{i,\alpha}$
variables are to be identified.
Second, continuity and dropping rules are to be explicitly taken into account.
To address these issues it is natural to introduce the notion of patch, which is the sequence of
several sequentially groupped connected plies of the same area shared between different zones, Fig.~\ref{fig::patch}. 
Note that patch concept had been considered in the literature already~\cite{MILP2,graph1,graph2},
albeit with different purposes.
Every patch is characterized only by its width and the connected set of zones it covers (patches covering all zones
will be refered to as having no boundary or continuous).
Upon stacking sequences determination one could also assign a particular ordered ply orientations to every patch.
Description of multi-zoned structure in term of patches allows to address directly both continuity and dropping
rules, as well as to properly assign the above binary degrees of freedom to different zones.
Indeed, continuity rule only requires that boundaries of different patches never come to the same location.
Dropping rule states that width of patches with non-zero boundary is limited from above.
Therefore, the continuity and dropping rules for stacking sequences are transferred identically
to the particular requirements on admissible patches.

When appropriate exhaustive set of patches is constructed, stacking sequence determination might be
performed similar to single zone case. Distinctions are mostly technical, although efficient solution requires
specific tricks (see below). For instance, binary variables and corresponding
``horizontal'' constraints are to be considered now for plies in all patches,
while ``vertical'' constraints are to be introduced for each zone with respect to all covering patches.
With these reservations contiguity and disorientation
requirements look almost identical to what had been discussed already.

Therefore, the prime difficulty to construct stacking sequences for multiple zones reduces
to finding an exhaustive set of patches compatible with continuity and dropping requirements at zone boundaries
and to the solution of resulting linear integer problem.
These are the topics of next Section.

\section{Stacking Sequences Reconstruction}
\label{section::SLB}
In this section we provide the detailed description of computationally inexpensive method to
reconstruct full set of stacking sequences of multi-zoned laminate composite structures.
The problem setup is as follows. There is a known number of zones $Z$ of fixed connectivity,
for which both the target number of plies $n_{\mu,\alpha}$ $\mu = 1,..., Z$ and the desired lamination
parameters values $\xi_{\mu,\alpha}$ are given. Our method consists of two prime stages:
\begin{enumerate}
\item selection of exhaustive set of patches $p$, fulfilling continuity and dropping rules;
\item two-step solution of resulting linear integer problem aimed to ensure contiguity and
  disorientation requirements and to provide minimal $L_1$-residual with respect to
  externally provided zone signatures $n_{\mu,\alpha}$ and  lamination parameters $\xi_{\mu,\alpha}$,
  $\mu = 1,..., Z$.
\end{enumerate}

\begin{figure}[t]
\centerline{\includegraphics[width=0.5\textwidth]{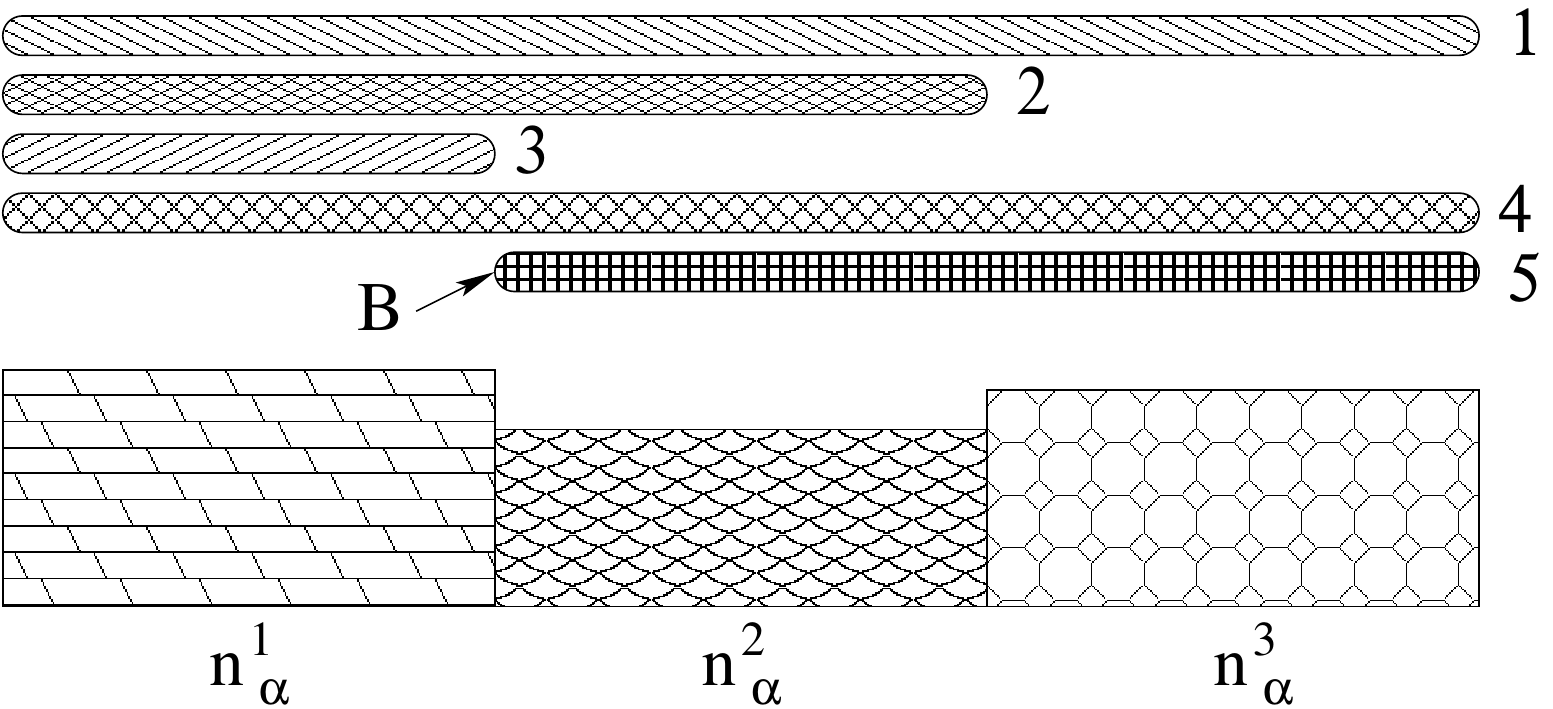}}
\caption{Particular iteration of SLB-like methodology in 3-zone example. Five patches already constructed,
indices indicate construction order. Boundary set $B$ is non-zero, $n^\mu_\alpha$ are the reduced zones signatures
to be exhausted during the process.}
\label{fig::slb1}
\end{figure}

The first stage of the method is reminiscent to well-known Shared Layer Blending (SLB)
methodology~\cite{toropov1,toropov2,Liu-phd} and
at each step attempts to construct patches of maximal permitted area to be placed 
most distantly from the mid-plane (outer blending). At every iteration
of the algorithm the following data is available, Fig.~\ref{fig::slb1}:
\begin{itemize}
\item Current signatures of zones $n_{\mu,\alpha}$;
\item Boundary set $B$  of already constructed patches, which are the locations of ply drops
  as they are seen from the mid-plane.
\end{itemize}
To obtain the next patch we sequentially consider all possible integer vectors $t_\alpha$ such that
at least for one zone $t_\alpha \le n_{\mu,\alpha}$, $\forall \alpha$.
For given signature $t$ patch candidate is constructed similar to that in conventional SLB,
however, we attempt to get all possible candidates.
Moreover, it is crucial that not all formally acceptable vectors $t$ are indeed appropriate.
One should check first that accumulated patch set together with 
proposed candidate and remaining zones signatures
leaves a chance for resulting linear integer problem to be feasible.
Of course, it is not technically possible to solve full fledged integer linear problem for each patch candidate and
we utilize simplified scheme, in which feasibility is tested with respect to corresponding relaxed variables.
Namely, we augment accumulated patch stack with candidate patch and
patches of width $\sum_\alpha n_{\mu,\alpha} > 0$,
each covering one remaining zone (thus, the rest of blending rules is temporally discarded)
and attempt to solve linear problem of type \ref{linear-problem1} (see the previous Section)
treating all $s_{\mu,\alpha}$ variables as continuous.
This provides crude but rather restrictive feasiliblity test and ensures that
proposed candidate makes sense at all.
In practice, feasibility test discards an absolute majority of proposed patches and here we radically
differ from conventional SLB.
Collected set of patch candidates are then sorted with respect to area and width, most extended and thick
candidates are to be considered first.
In case current boundary set $B$ is non-zero or patch candidate is not continuous,
it is also confronted to continuity and dropping rules: candidates violating any of them are either modified
(diminished in width and/or area) or discarded.
First encountered suitable candidate is taken to be the next patch.

One additonal point is worth mentioning here.
It might happen, especially at later iterations, that none of candidates pass feasibiblity check.
There might be several possible causes: either boundary set $B$ prevents patch selection or relaxed
linear problem appears to be infeasible. However, it is crucial that in both cases one and the same trick,
inspired by conventional SLB methodology, recovers normal patch selection algorithm.
Namely, it suffices to borrow one continuous ply from some already
constructed patch (perhaps, outer most one) and place it at current position in the patch stack.
Ultimate reason for this recipe to succeed is that it simultaneously removes boundary set $B$ altogether,
diminishes the number of disorientation interfaces and introduces additional layer
in the bulk, which might help to satisfy contiguity rules.
The only situation, when this trick might not be applicable, corresponds to already exhausted outer-most patch,
but this is virtually impossible in practical applications.
In these rare cases we simply bypass feasibility checks, patch construction
is performed similar to that in conventional SLB methodology.
Once appropriate patch with signature $t$ is selected, covered zones are reduced
$n_\mu \gets n_\mu - t$ and process repeats until signatures are zeroed in all zones.

\begin{figure}[t]
\centerline{\includegraphics[width=1.\textwidth]{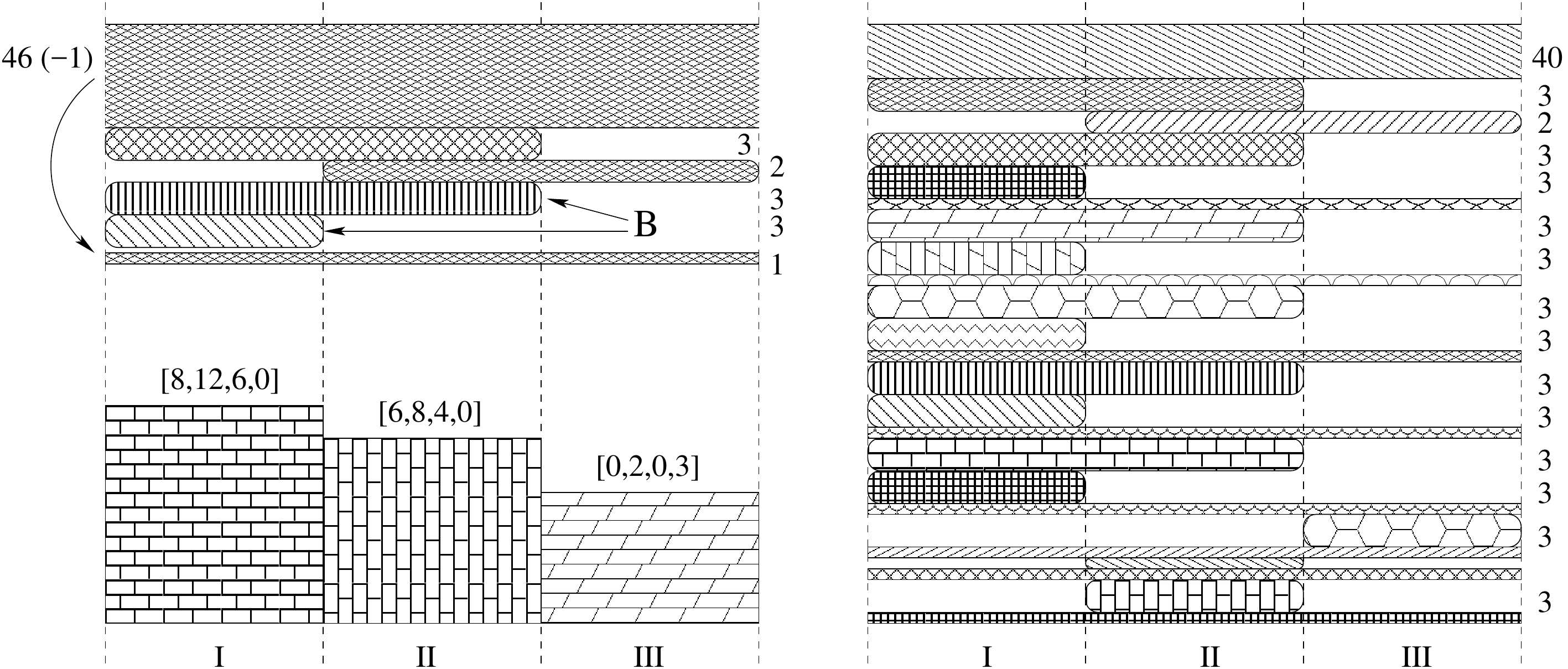}}
\caption{Patch construction algorithm in the example of three line-aligned zones.
Left: intermediate state, some patches are completed, non-empty boundary requires
to borrow one ply from outer-most layer; reduced zones content is shown at the bottom.
Right: final exhaustive set of patches in all zones. Nearby numbers indicate patch widths. }
\label{fig::slb-example}
\end{figure}

For illustration purposes we follow Ref.~\cite{toropov1} and
consider the above procedure in simple case of three neighboring
line-aligned zones $1-2-3$ with initial signatures $[40/17/7]_s$, $[35/14/9]_s$ and $[29/6/12]_s$,
Fig.~\ref{fig::slb-example}.
Algorithm first selects appropriate outer-most patch of width 46 and signature $[6,27,6,7]$, which is written
in general form $[n_{-45}, n_0, n_{45}, n_{90}]$ because each particular patch is not required to be symmetric.
Implied feasibility checks make our method different from conventional SLB at first step already,
SLB would pick up outer layer with signature $[6,29,6,7]$ and width 48, but it does not pass consistency
test in our approach.
After appropriate reductions algorithm continues and second suitable patch is determined
to be $[1,0,2,0]$, it covers only first two zones. Non-empty boundary set between second and third zones
appear, it has to be taken into account in subsequent construction.
Third found patch is again not continuous, has signature $[0,0,0,2]$ and covers second and third zones,
boundary remains non-empty, but it is now shifted to be inbetween first and second zones.
Fortunately, boundary does not pose serious obstacles for next two iterations of the algorithm,
at which patch set is augmented with $[1,0,2,0]$ (first two zones) and $[1,1,1,0]$ (first zone only).
At this stage (Fig.~\ref{fig::slb-example}, left)
boundary set becomes maximal and the only possibility to continue is to borrow several plies from outer-most patch.
It turns out that single continuous ply placed at current location passes feasibility check. Algorithm in a sense
is restarted at this point because boundary set is totally removed.
Final patch structure is presented on the right panel of Fig.~\ref{fig::slb-example},
where the relative widths of all but outer-most patches are correctly preserved.
Once the exhaustive patch structure is determined, stacking sequences are reconstructed as described in previous Section.
Note that we solve symmetric with respect to $\rho$, $\xi$ variables formulation, therefore solution
is not obliged to reproduce exactly initial ply counts in the zones.
In particular, in the considered example output zones signatures are $[39/17/8]_s$, $[35/14/9]_s$ and $[28/8/9]_s$ respectively
with the same total number of plies, but slightly (of order $1/N$) different percentages.

The outcome of the above procedure is the ordered set of patches, each characterized by its width and
covered set of zones.
Then determination of stacking sequences requires to solve integer linear problem,
single-zone version of which  was described in the previous Section.
For reasons already explained we solve symmetric problem, in which percentages $\rho$ and
lamination parameters $\xi$ are treated on equal footing.

It is apparent that for sufficiently fine zoning the linear integer problem might become prohibitively large,
especially when zone signatures are vastly different and corresponding patch set contains significant number
of small patches.
Then it was proven advantageous to conduct the solution via the two stage process.
First, stacking sequence is determined for outer-most continuous patch.
Then reduced linear integer problem is solved to establish ply orientations in remaining patches.
Underlying idea is essentially the same as in SLB method: in practically all relevant applications
top level continuous patch is expected to contain significant number of plies.
However, the treatment of outer most patch is much simpler and is equivalent to single zone stacking sequence determination.
Therefore, two-stage procedure allows to greatly reduce the overall complexity of the algorithm.

It remains to discuss how algorithm recovers in rare cases of final integer linear problem infeasibility.
Since continuity and dropping rules are identically fulfilled, binding constraints, which might prevent the solution,
are related to either contiguity or disorientation rules.
In this case we essentially apply the same trick as above, however, appropriate places to insert continuous plies
are to be determined first. In order to establish proper locations we solve specifically
reformulated integer linear problem, in which all potentially binding constraints
are relaxed and objective function is $L_1$ norm of corresponding relaxation variables.
Reformulated problem is always feasible and its solution identifies the minimal set of
constraints preventing stacking sequences determination. Then we attempt to add a continuous plies
(perhaps, borrowed from outer-most patch) at several determined locations and solve original problem anew.
In case of failure algorithm terminates and stacking sequences remain undefined.
However, this happens too rarely to be of any practical relevance.

\section{Design Space}
\label{section::domain}

As was argued previously, retaining clear distinction between
macroscopic (plies count $N$, percentages $\rho$ and lamination parameters $\xi$) and
microscopic (specific sequence of ply orientations) ``degrees of freedom''
is ought to be the cornerstone of efficient optimization strategy.
However, integral characteristics appear as redundant variables, for which
proper variation ranges and constraints on allowed values are yet to be given.
This is the problem of integral parameters feasible domain determination,
which deserved much attention in the past~\cite{domain1,domain2,domain3}.

Multiple zones are of no concern here because total allowed design space is the appropriate
direct product of identical single zone feasible domains.
While there are essentially no specific limits imposed on the number of plies $N$,
it had been known long ago that once percentages $\rho$ are fixed, feasible region $\Psi_\rho(\xi)$ of 
lamination parameters $\xi$ is convex polyhedra, vertices of which are rather explicit geometrically.
Namely, consider an ``extreme'' configurations consisting of $N_\theta$ groups of sequentialy stacked
plies of the same orientation.
There are $N_\theta !$ stacking sequences of this type and they define the vertices
of $\Psi_\rho(\xi)$: convex hull of corresponding ``extreme'' $\xi$-values is
an exact feasible region for $\xi$ parameters at fixed $\rho$
\beq
\label{xi-linear1}
\Psi_\rho(\xi) ~=~ \{ \xi \,\, : \,\, A(\rho)\, \xi ~\le~ b(\rho) \, \}\,.
\eeq
This immediately suggests~\cite{self} to utilize nested collaborative methods to conduct optimization
with respect to $(N, \rho, \xi)$ parameters. Namely, valid direct methodology is to treat
$N,\rho$ parameters subject to
\beq
N ~\ge~ 0\,, \qquad  0 ~\le ~ \rho ~\le ~ 1\,, \quad \sum_i \rho_i = 1\,,
\eeq
at outer level of nested scheme, while lamination parameters $\xi \in [-1;1]^3$ together with
linear constraints~(\ref{xi-linear1}) are to be considered at inner level.
Advantage of this approach is that it exactly covers feasible region $\Omega$ of $(\rho,\xi)$ parameters.
However, it is not symmetric in $(\rho,\xi)$ and, as might be generically expected for collaborative
schemes, is less efficient in terms of required number of model evaluations compared to direct approaches.
Therefore, it is desirable to establish tight approximation to feasible domain in the combined 
$(\rho,\xi)$ space.

Our proposition is that feasible region $\Omega$ in the space of combined lamination parameters $(\rho,\xi)$
might be described rather precisely in the product form:
\beq
\label{feasible-domain-proposition}
\Omega ~\approx~ \Omega_\rho \, \times \, \Omega_\xi\,,
\eeq
where:
\begin{itemize}
\item $\Omega_\rho$ is convex polytope in $\rho$-subspace
  \beq
  \label{rho-limits}
  \Omega_\rho ~: \qquad \rho_i \,\in\, [\,\rho_{min}\,;\, 1 \,-\, (N_\theta - 1)\, \rho_{min}\,]\,, \quad \sum_i \rho_i = 1
  \eeq
  with $N_\theta$ vertices $\rho_v$, for which only one $\rho$-component takes maximal allowed value,
  indicated in (\ref{rho-limits}), others remain at minimal bound $\rho_{min}$;
\item $\Omega_\xi$ is again a convex polytope, constructed similar to $\Psi_\rho(\xi)$ above,
  but using $\rho_v$ as reference percentages
  \beq
  \Omega_\xi ~=~ \Psi_{\rho_v}(\xi)\,.
  \eeq
  Note that since $\Psi_\rho(\xi)$ explicitly accounts for all permulations of ``extreme'' stacking sequences,
  $\Omega_\xi$ is in fact $\rho_v$ independent.
\end{itemize}
In order to verify and access the quality of proposition (\ref{feasible-domain-proposition})
one should check the inclusions
\beq
\label{inclusion}
\Omega_\rho \, \times \, \Omega_\xi ~\subseteq ~ \Omega\,,
\quad \qquad
\Omega ~\subseteq ~ \Omega_\rho \, \times \, \Omega_\xi\,,
\eeq
and estimate the quantity
\beq
\label{Delta}
\Delta ~=~ |\,(\Omega_\rho \, \times \, \Omega_\xi) \, \backslash \,  \Omega\,| \,/\, |\Omega| \,,
\eeq
which is the relative difference of volumes of two sets (\ref{feasible-domain-proposition}).
To this end we consider the probabilities
\beq
\label{prob1}
P[ x \notin \Omega_\rho \, \times \, \Omega_\xi] \qquad \mbox{for} \qquad x ~\sim ~ U[\Omega]\,,
\eeq
\beq
\label{prob2}
P[ x \notin \Omega] \qquad \mbox{for} \qquad x ~\sim ~ U[\Omega_\rho \, \times \, \Omega_\xi]\,,
\eeq
where $U[W]$ denotes uniform distribution over the set $W$.
Note that (\ref{prob1}) directly measures the relative volume difference $\Delta$, while (\ref{prob2}) provides
the same with $\Omega$ and $\Omega_\rho \, \times \, \Omega_\xi$ interchanged.

Probabilities (\ref{prob1}, \ref{prob2}) were measured via Monte-Carlo methodology using
sufficiently large ($10^8$) sample of uniformly distributed points.
It turns out that probability (\ref{prob1}) is zero with high confidence level
(all points, sampled from $U[\Omega]$, belong to $\Omega_\rho \, \times \, \Omega_\xi$)
\beq
\Omega ~\subset ~ \Omega_\rho \, \times \, \Omega_\xi\,,
\eeq
while (\ref{prob2}) is non-zero but small providing the following estimate of relative volume difference
\beq
0 ~< ~ \Delta ~\lesssim ~ 10^{-5}\,.
\eeq
Therefore, we heuristically justified the proposition (\ref{feasible-domain-proposition}). 
Moreover, direct product $\Omega_\rho \, \times \, \Omega_\xi$ provides conservative estimate for true
feasible region $\Omega$ and is sufficiently tight for any practical purposes.
Simple to describe approximation $\Omega_\rho \, \times \, \Omega_\xi$ 
provides the base for efficient optimization methodology to be described below.

\section{Hierarchical Zoning}
\label{section::zoning}
Hierarchical zoning alluded to in Section~\ref{section::intro} is an important ingredient
of our optimization methodology and is worth of separate discussion.
Indeed, in industrial applications it is not a rare case to consider sufficiently large
number of zones (up to a few decades).
The number of design variables might then becomes as large as $O(100)$, which is about or
even beyond the upper limit for engineering optimization methods.
However, the challenge here is not only technical: the prime point is that blind application
of even most powerfull optimization algorithms is not the most efficient approach to physically
motivated problems.
Indeed, appropriate solution is known to be smooth in a sense that
it possesses some finite length scale of characteristic variations.
However, when number of introduced variables becomes too large compared to that length scale
optimization process will definitely stall: just by entropy arguments any algorithm will spend almost
all the efforts in probing physically irrelevant configurations, corresponding to rough distributions.

Unfortunately, it is usually not possible to guess the proper length scale {\it a priori},
it is known to be too much case to case dependent.
Moreover, some bounds on it might come from external arguments, like ribs/stringers locations etc.
In either case, guessing the length scale is known to be error-prone and too much sensitive to various problem details.
Instead, another approach, known as multi-scale analysis,
was proven to be successful in numerous applications: one starts with largest
available length scale (coarse resolution, small number of variables), finds the approximate solution 
and then dimishes the length scale (refines the resolution, introduces more variables) using previous
solution as the initial guess for the next iteration.
Multi-scale methodology is quite generic and works in virtually all cases, however, its silent feature
to be explicitly mentioned here is that it assumes the use of localized optimization algorithms.
Indeed, the very notion of ``initial guess'' makes sense only in local settings and becomes
void once global search is attempted.
We will come back to this issue when discussing our optimization methodology.

\begin{figure}[t]
\centerline{\includegraphics[width=0.75\textwidth]{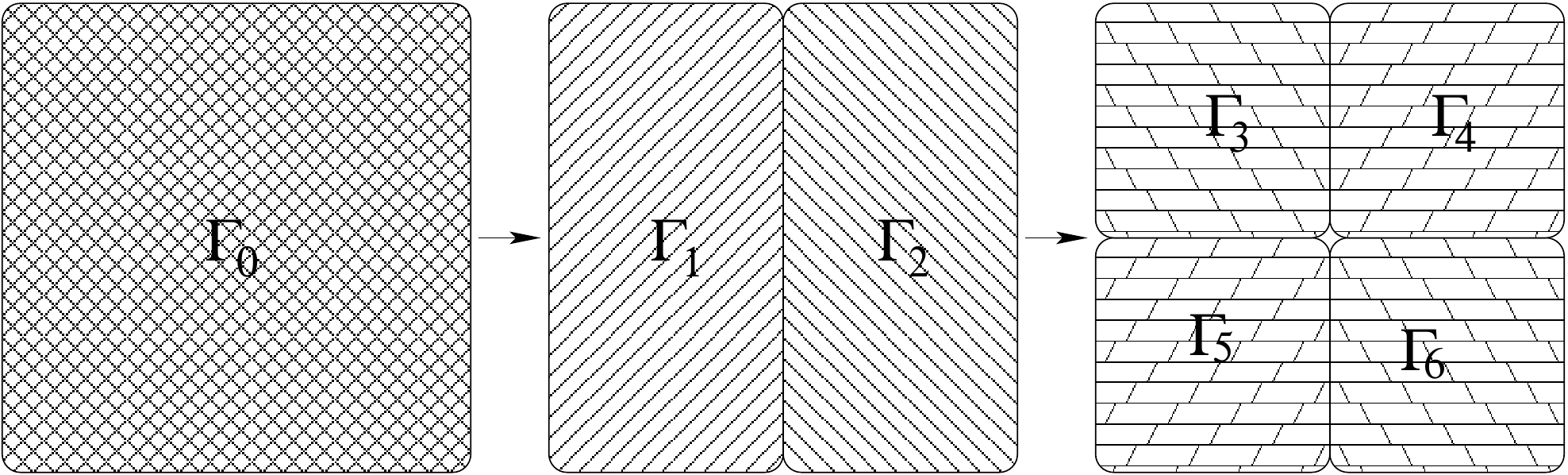}}
\caption{Example of three-level hierarchical zoning.}
\label{fig::3-zoning}
\end{figure}

Within the context of laminated composites optimization multi-scale methodology
looks extremely simple and reduces to sequential refinement of nested zones
within the original structure geometry $\Gamma$.
Indeed, the ultimate goal of zoning procedure is to find subdivision of original geometry into smaller
components (zones) $\Gamma_\mu$
\beq
\label{zone-union}
\Gamma ~=~ \cup_\mu \Gamma_\mu\,,
\eeq
which is sufficiently effective in total mass reduction, but yet is technically feasible to achieve.
Of course, there is an infinite number of ways how actual division might be performed, however,
several competitive factors are to be taken into account.
On one hand, smaller total mass requires larger number of zones. 
On the other hand, gain in mass diminishes with each resolution refinement, while
finding the solution becomes more and more expensive.
Usually there are some technological constraints on smallest possible zone geometry
and it is not a rare case when finest reachable zoning structure is known in advance.
Although theoretically it is possible to discuss fully automated zoning, we are
currently not in a position to do this quantitatively.
Instead, we assume that smallest possible zones geometry $\Gamma_\mu$ is known beforehand and
solve the problem how to obtain corresponding optimal parameters $(N,\rho,\xi)_\mu$
adaptively. Schematically, utilized procedure looks as follows.
We always start with most coarse zoning, in which the whole structure is considered as a single zone, $\Gamma^{(0)} = \Gamma$.
Corresponding optimal solution is then easy to obtain, although it might be far from real optimality.
At every iteration we have determined therefore some exhaustive set of zones $\Gamma_\mu$,
fulfilling (\ref{zone-union}) together with corresponding zone-specific parameters $(N,\rho,\xi)_\mu$.
Then some of $\Gamma_\mu$ are divided further into several smaller zones and localized optimization
method (see below) is applied to get parameters of smaller zones using previous solution as the initial guess.
Process repeats until either smallest allowed zones geometries are reached or the gain in total mass becomes
negligible. Three-step zoning procedure is illustarted on Fig.~\ref{fig::3-zoning}.
It should be noted that adaptive choice of zones is only the convenient representation of entire
composite structure during the optimization process.
Any involved modeling tools always operate in terms of smallest available zones regardless
of current zoning level.

\section{Optimization Methodology}
\label{section::method}
This section provides a synthesis to what had been discussed so far
and assembles into efficient optimization methodology prime ingredients of our approach:
hierarchical zoning, separation of macro- and micro-scopic scales,
computationally inexpensive (``on-line'') reconstruction of stacking sequences fulfilling
all the required blending rules.
We consider the following generic setup.
There is a composite structure of fixed geometry $\Gamma$, for which 
finest allowed subdivision into elementary subzones is provided,
$\Gamma = \cup_{k \in K} \Gamma^{(0)}_k$ (from now on elementary zones, which cannot be refined further
and for which the geometries $\Gamma^{(0)}_k$ are assumed to be known, are refered to as subzones).
Zones are then defined as non-intersecting unions of several connected subzones
\beq
\Gamma_\mu ~=~ \cup_{k \in K_\mu} \Gamma^{(0)}_k\,, \qquad
\cup_\mu K_\mu ~=~ K\,, \qquad
K_\mu \cap K_\nu ~=~ \oslash\,,
\eeq
connectivity structure of which could be derived from that of subzones.
The purpose is to find reasonably efficient zoning $\Gamma = \cup_\mu \Gamma_\mu$
and stacking sequences $S_\mu$ in all zones, for which total mass of the structure is minimal
provided that specific set of reserve factors $\rf_k$, associated with each elementary subzone,
remains sufficiently large, $\rf_k \ge 1$.
Without loss of generality we assume the same normalization of all reserve factors, which allows us
to define zone-specific single RF-function:
\beq
\label{zone-rf}
\rf_\mu ~=~ \min\limits_{k \in K_\mu} \,\, \rf_k
\eeq
It is assumed that each reserve factor $\rf_k$ is to be estimated with some externally given
modeling software operating in terms of stacking sequences $S_k$, which for selected zoning
are the same within the particular zone, $S_k = S_\mu$, $k \in K_\mu$.

Proposed methodology is iterative and as follows from above generic setup its outer-most
level is ought to be hierarchical zoning refinement. Initially we consider trivial partitioning
consisting of single zone. Then at every iteration current zones are divided into
two or three smaller zones until either minimal geometries $\Gamma^{(0)}$ are reached
or termination criteria trigger algorithm completion.
Therefore at every outer-most step the following data is available:
\begin{itemize}
\item definition of zones $\{K_\mu\}$ as the collections of connected elementary subzones with empty pair-wise intersections;
\item current approximations $X^*_\mu \equiv (N^*,\rho^*,\xi^*)_\mu$ to optimal solution in each zone.
\end{itemize}
Inner loop of our algorithm performs constrained minimization of total mass
\beq
\label{opt-problem}
\begin{array}{c}
\min_X \, \sum_\mu \, M(X_\mu) \\
~ \\
\rf_\mu(X) ~ \ge ~ 1
\end{array}
\eeq
with respect to the design variables
\beq
\label{total-design-vars}
X ~=~ [ \, X_1, \, ... \, , X_Z \, ]\,,
\eeq
which are the collection of number of plies $N_\mu$, percentages $\rho_\mu$ and lamination parameters $\xi_\mu$
for all current zones.
Variation domain of design variables is the direct product of feasible regions, determined for each factor
$X_\mu$ separately as described in Section~\ref{section::domain}.

As we already noted direct optimization of (\ref{opt-problem})
is not only problematic technically, but is also not the most efficient approach.
Indeed, the problem (\ref{opt-problem}) possesses specific structure, namely,
various factors $X_\mu$ are totally factorized
in the objective function and their interdependencies are only due to the imposed constraints.
Furthermore, in all physically motivated problems some sort of locality is expected to hold.
Qualitatively, altering the stacking sequence $X_\mu$ in a particular zone mostly affects
reserve factors of this and some finite number of neighboring zones, while too distant $\rf_\nu$
remain practically the same. Quantitatively, this means that in the decomposition
\beq
\delta \rf_\mu \,/\,\delta X_\nu ~=~ D \cdot \delta_{\mu\nu} ~+~ Q_{\mu\nu}
\eeq
diagonal term $D$ dominates, while off-diagonal entries $Q$ rapidly fall down with the distance
between zones $\mu$ and $\nu$.
Practically, however, it is not possible to infer appropriate length scale, which dictates the number
and locations of dependent neighbors.
Moreover, situation often becomes more complicated when additional structural elements
(beams, ribs and such), not present explicitly in problem formulation, mediate the interactions.
Given that we decline to determine zone interdependencies in advance, instead we
propose to do that adaptively with coordinate-descent like optimization method,
in which dependencies are estimated dynamically during the process.

In more details, basic inner loop of our algorithm considers design variables $X_\mu$ sequentially
(after appropriate ordering, see below). At iteration $\mu$ optimization is performed in
the reduced design space
\beq
\label{reduced-design-vars}
X ~\in~ [ \, X^*_1, \, ... \, , X^*_\mu \,\pm\, \Delta\,,\, ... \, , X^*_Z \, ]
\eeq
with respect to $X_\mu$ variables only, subject to linear constraints defining the appropriate
feasible region.
In other words, all the variables are held fixed at current guess $X^*_\nu$,
except $X_\mu$, which is allowed to vary within the adjustable region around the current solution.
Note that despite of the reduction of the design space the number of constraints does not change,
at every iteration all zone-specific reserve factors (\ref{zone-rf}) are taken into account.
Particular feature of our methodology is that currently known best solution explicitly enters
the formalism, which therefore becomes directly applicable in multi-scale context (see above).

However, there remain two crucial issues yet to be resolved.
First, it is well known that coordinate descent methods applied in constrained problems
often get stuck in even not a locally optimal locations, because, in particular, coordinate axes
have nothing to do and are not aligned with tangent space to currently active constraints.
Second, due to the usual computational complexity of engineering applications it is custom to use
Surrogate-Based Optimization (SBO) methods to search for the solution.
However, SBO by its nature is a global methodology, for which the notion of proper
initial guess is rather illusive.

To circumvent both issues we propose to supplement SBO methodology, applied in the reduced
design space (\ref{reduced-design-vars}), with filtering approach, well known in mathematical
programming (see, e.g., Refs.~\cite{filter1,filter2,filter3} and references therein).
In a nutshell, filter is a binary marker, which operates in two-dimensional space
of objective function values and single measure of constraints violation and marks every evaluated
design as being ``good'' if it improves currently selected Pareto optimal set of solutions.
Most important for us feature of filtering is that it might allow and mark as acceptable
the designs with rather significant constraints violation.
This suggests that filter might be an invaluable tool within the considered methodology.
In more details, we switch on the filter facility at the beginning of every cycle of inner optimization loop.
Then the next SBO optimization with respect to $X_\mu$ variables is peformed
as usual, however, the selection of best found solution is done only among the designs accepted by filter.
In particular, best accepted solution becomes the ``base point'' $X^*$ for the next SBO iteration dealing
with next group of variables $X_\nu$.

Experience revealed that filtering almost totally ameliorates
the deficiencies of utilized SBO-based coordinate descent method.
However, to be on the safe side we also monitor interdependencies of various groups
of variables and attempt to perform optimization of several zones simultaneously
in case no progress had been achieved in last performed inner loop
with feasible base point $X^*$.
Indeed, coordinate descent with respect to $X_\mu$ might get stuck due to the activity
of some constraints $\rf_\nu$, variation of which with $X_\mu$ precludes the progress of the algorithm.
Within the SBO context $X_\mu$ variables are sampled somehow and it is natural then to consider
violation probabilities of different constraints.
Zones with maximal violation probabilities are dependent with the considered one and
most likely prevent algorithm progress.
Due to the mentioned above locality properties the number dependent active zones is usually small.

\begin{algorithm}
  \DontPrintSemicolon
  \KwData{zoning structure $L$}
  \For{$\forall$ zoning levels $l \in L$ }{
    $X^* ~ \gets ~ $ best solution\;
    reset filter\;
    \For{$\forall$ zones $\mu \in l$ }{
      optimize for $X_\nu \in X^*_\nu \pm \delta_{\mu\nu} \, \Delta$, $\quad \forall \nu$\;
      collect dependencies $D_\mu = \{\mu, \mu_1, \mu_2,...\}$\;
      $X^* ~ \gets ~ $ best accepted by filter\;
    }
    \If{ solution not improved }{
      \For{$\forall$ zones $\mu \in l$ }{
        optimize for $X_\nu \in X^*_\nu \pm \Delta \cdot \left\{\begin{array}{l} 1, \,\, \nu \in D_\mu \\ 0, \,\, \nu \notin D_\mu\end{array}\right.$,
        $\quad \forall \nu$\;
      }
    }
    \If{ termination criteria }{
      break
    }
  }
 \caption{Summary of implemented optimization scheme.}
 \label{fig::opt-scheme}
\end{algorithm}

To complete the specification of our optimization methodology it remains to specify termination criteria
and discuss the best suited ordering of zones to be considered in inner optimization loop.
Termination criteria are simple: either no progress was achieved during the last cycle or
current best solution is feasible with all the constraints being active.
As far as ordering is concerned, we found that it is advantageous to consider first
infeasible zones with $\rf_\mu < 1$.
Then inactive constraints with corresponding groups of variables $X_\mu$ are worth to treat next.
Finally, active zones are to be optimized, but with one essential modification.
For active zones there is no hope to improve the mass. Therefore, it makes sense
to fix the number of plies $N_\mu$ for active zone and to maximize its reserve factor with respect
to remaining variables $(\rho,\xi)_\mu$.
This is essentially equivalent to the selection of best stacking sequence at fixed total mass.

Simplified scheme of our optimization algorithm is presented on Fig.~\ref{fig::opt-scheme}.
Note that we do not discuss the details of utilized SBO methods, these were
provided by algorithmic core of \pS integration platform~\cite{DATADVANCE},
on which our technical implementation is based.
It is important and very convenient that stacking sequence reconstruction remains an optional
feature in the above scheme.
Depending on specific modeling software one could use either full-fledged optimization
with ``on-line'' determination of allowed stacking sequences or fall back to conventional bi-level strategy. 

\section{Application Examples}

In this section we discuss two application examples, with which we tested our methodology.
Tests selection is primarily dictated by availability of corresponding technical details in the literature,
unfortunately, for many interesting applications technicalities could not be extracted from public sources.
First test case is a well known horseshoe problem~\cite{blending1,horseshoe1},
in which zone loads are fixed and no issues with zones interdependencies arise.
Historically, we considered it at early stages of reported study using yet unfinished versions of our methodology.
For that reason we believe that presented results are not the best attainable, however,
they are quite satisfactory and hence deserve some discussions.
Second example is a conventional wing box test problem~\cite{toropov1,toropov2,Liu-phd,wb1}, which adequately reflects
majority of issues present on real-life applications: loads redistribution upon the change of
single zone parameters, presence of ``hidden'' structural elements etc. 
It is important that both examples might be considered with no explicit reference
to specific stacking sequences (without blending rules applied).
Corresponding treatment is equivalent to the solution of upper level problem in conventional
bi-level scheme and provides strict lower bound on attainable total mass, which is to be
compared with the solution respecting blending rules.
In both cases we found that reconstruction of proper stacking sequences worsens the solution as it should,
but optimal mass remains within 10\% gap from corresponding theoretical lower bound.

\subsection{Horseshoe Model}
\label{section::hs}

\begin{figure}[t]
\centerline{\includegraphics[width=0.75\textwidth]{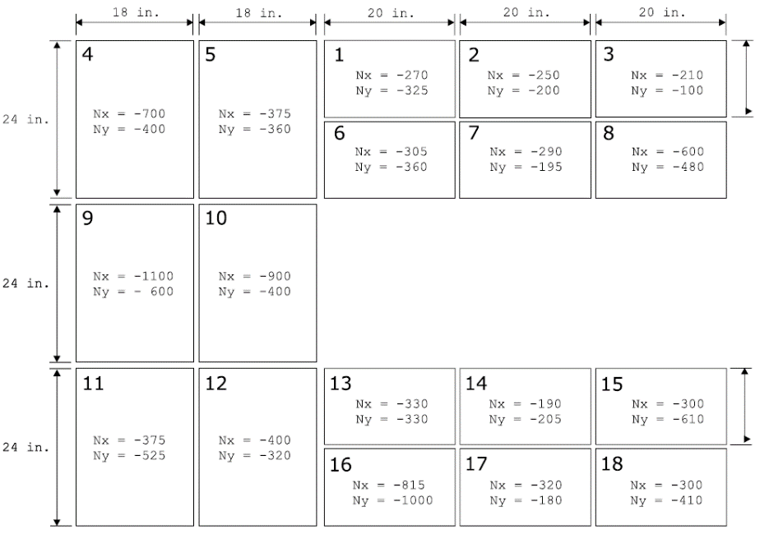}}
\caption{Horseshoe test case setup: geometry and applied fixed loads.}
\label{fig::hs-general}
\end{figure}

Setup of horseshoe problem is illustrated on Fig.~\ref{fig::hs-general}.
Plain horseshoe-like structure consists of 18 elementary subzones, for which the geometry and applied loads are known in advance.
Objective function is the total mass, while 18 constraints correspond to the stability requirements
for all subzones (buckling reserve factors).
Since there are no shear stresses buckling reserves might be evaluated semi-analytically~\cite{wb1}.

\begin{table}
\begin{center}
\begin{tabular}{|c|c|c|c|c|c|c|c|c|c|} \hline
      & 1   & 2   & 3   & 4   & 5   & 6   & 7   & 8   & 9  \\ \hline
N     & 11  & 9   & 8   & 16  & 14  & 12  & 10  & 13  & 20 \\ \hline
m     & 2.1 & 1.9 & 1.6 & 6.0 & 5.2 & 2.2 & 1.9 & 2.5 & 6.9 \\ \hline
$\rf$ & 1.0 & 1.0 & 1.0 & 1.0 & 1.0 & 1.1 & 1.0 & 1.0 & 1.0 \\ \hline
\end{tabular}\\
\begin{tabular}{|c|c|c|c|c|c|c|c|c|c|} \hline
      & 10  & 11  & 12  & 13  & 14  & 15  & 16  & 17  & 18 \\ \hline
N     & 17  & 14  & 14  & 11  & 10  & 13  & 14  & 9   & 12  \\ \hline
m     & 6.4 & 5.4 & 5.1 & 2.1 & 1.9 & 2.5 & 3.0 & 1.9 & 1.9 \\ \hline
$\rf$ & 1.0 & 1.0 & 1.0 & 1.0 & 1.0 & 1.0 & 1.0 & 1.0 & 1.1 \\ \hline
\end{tabular}
\end{center}
\caption{Horseshoe test case: total number of plies, masses [lb] and reserve factors
         obtained without blending rules imposition.}
\label{table::hs-theor}
\end{table}

First we consider the horseshoe test case without blending rules applied. 
Due to the fixed loads problem factorizes into 18 independent subproblems,
solution of which is straightforward.
Technically we performed optimization in \pS integration platform, which provides among other things efficient
optimization algorithms. Resulting optimal mass distribution in different subzones, corresponding reserve factors
and optimal number of plies in each subzone are summarized in Table~\ref{table::hs-theor}, where
we rounded all floats to one significant digit.
Conducted solution provides strict lower bound on the attainable mass
\beq
\label{hs-mass-th}
M^*_{th} ~=~ 61.0~\mbox{lb}.
\eeq
Once the solution at upper level of conventional bi-level scheme is obtained, it is instructive to consider
what happens if we solve lower level problem as well.
Direct application of stacking sequences reconstruction algorithm of Section~\ref{section::SLB}
reveals that in account of blending rules traditional bi-level scheme produces 18\% worse mass
\beq
\label{hs-mass-bi}
M^*_{bi-level} ~=~ 72.2~\mbox{lb}.
\eeq
Note that we intentionally do not provide the details of obtained solution.
As will become clear shortly it is not the best possible and for brevity reasons
corresponding stacking sequences and $\rf$ distribution are omitted.

\begin{figure}[t]
\centerline{\includegraphics[width=0.55\textwidth]{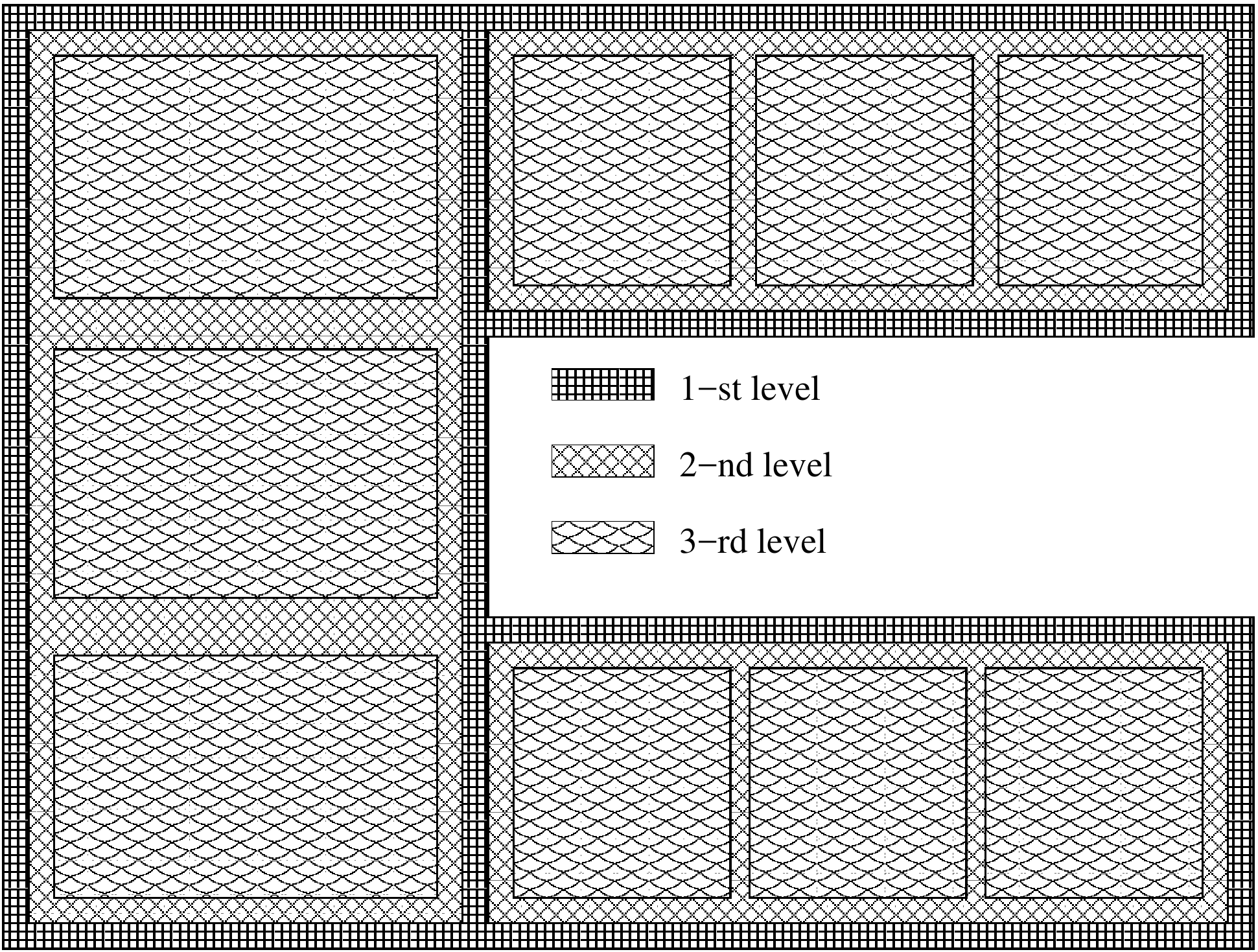}}
\caption{Hierarchical zoning in horseshoe test case: first three levels include 1, 3, and 9 zones,
last one consists of 18 zones, which are the same as elementary subzones.
For all but last levels zones ordering is counter-clockwise.}
\label{fig::hs-zoning}
\end{figure}

~

\begin{figure}[h!]
\centerline{\includegraphics[width=0.75\textwidth]{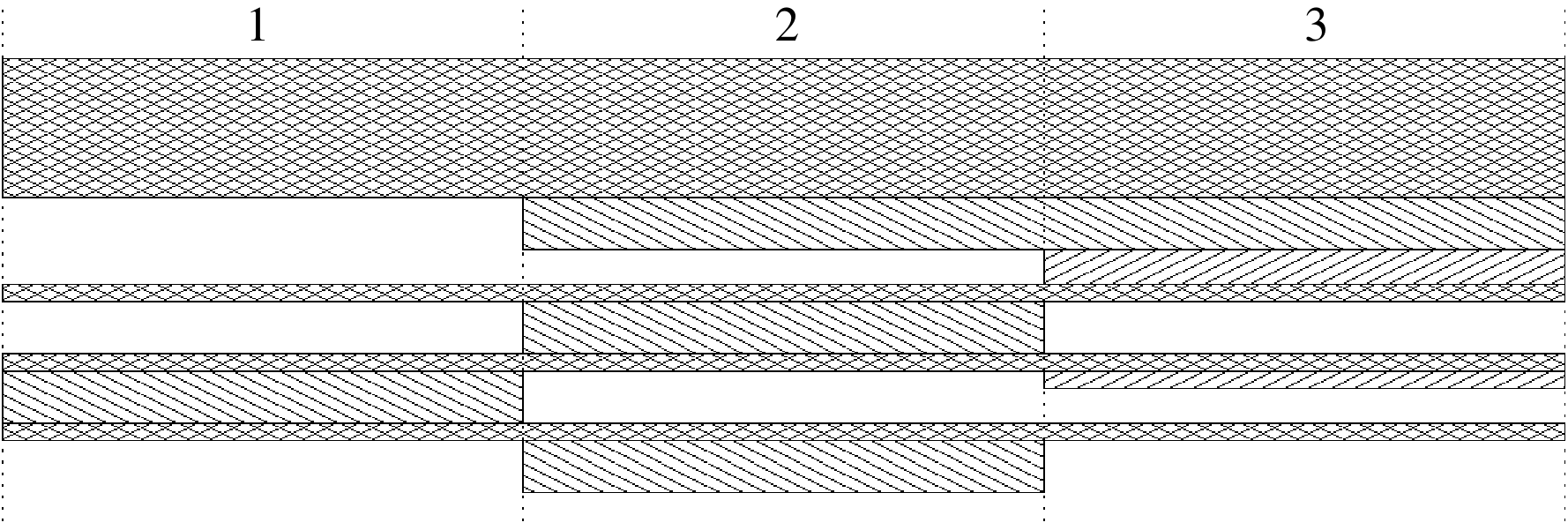}}
\caption{Horseshoe test case: patch structure of optimal solution at second zoning level.}
\label{fig::hs-patches-level2}
\end{figure}

~

\begin{figure}[h!]
\centerline{\includegraphics[width=0.75\textwidth]{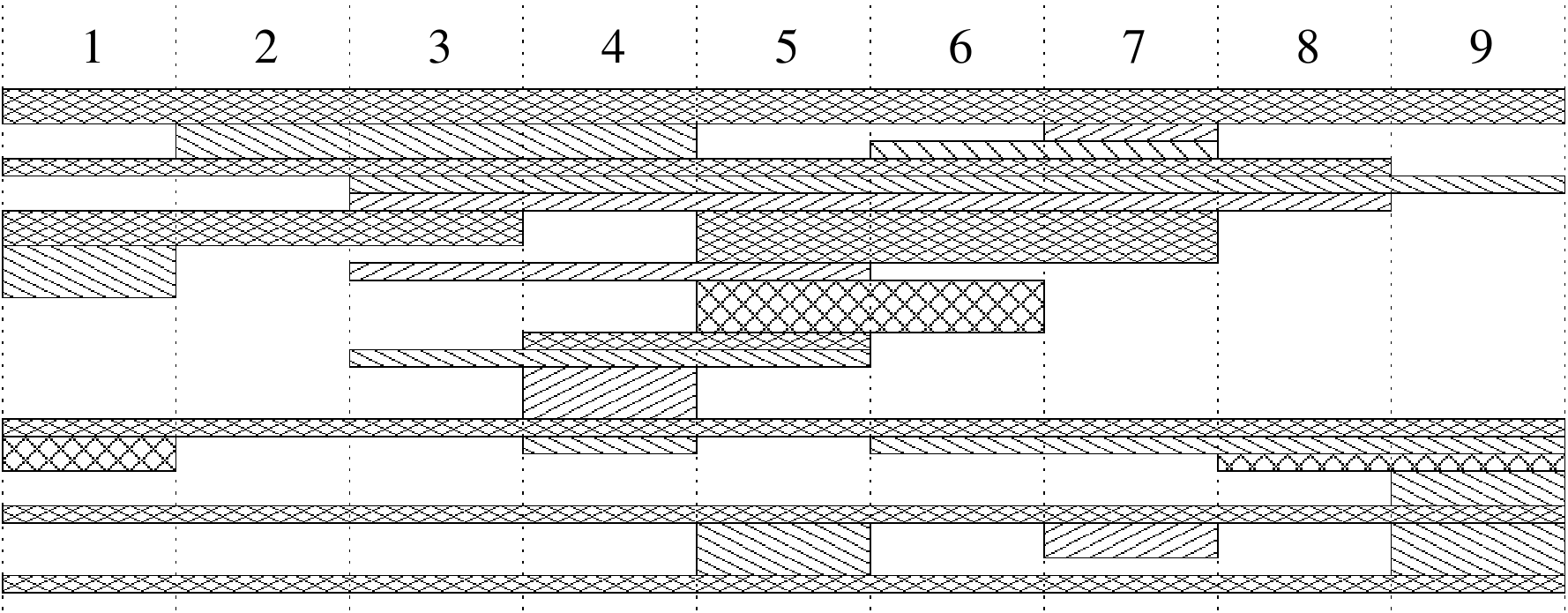}}
\caption{Horseshoe test case: patch structure of optimal solution at third (finest) zoning level.}
\label{fig::hs-patches-level3}
\end{figure}

Next let us consider the problem using methodology proposed in this paper.
First, we define the structure of hierarchical zoning: there are 4 zoning levels having 1, 3, 9, 18 zones,
respectively (Fig.~\ref{fig::hs-zoning}),
zones at last level coincide with elementary subzones on Fig.~\ref{fig::hs-general}.
At first zoning level the entire horseshoe structure is considered as a single zone, corresponding
solution does not involve patch construction algorithm.
Optimal stacking sequence at first level is
\beq
\label{hs-level1}
\begin{array}{c}
[ \, 8 \, / \, 4 \, / \, 4 \,]^{(1)}_s \\

[\mns{45}/0_2/\mns{45}/0_2/45/0/45/90_2/45/90/\mns{45}/90/\mns{45}/0_3/45]^{(1)}_s
\end{array}
\eeq
where conventional notations $[n_0/n_{\pm 45}/n_{90}]$ had been used and subscript indicates,
that this is only the half of total ply counts due to the symmetry (stacking sequences
are always ordered from mid-plane to outer skin). Corresponding mass and buckling reserve factor are
\beq
M^{(1)} ~=~ 93.6~\mbox{lb}, \qquad RF^{(1)} = 1.02\,.
\eeq

The above solution is used as an initial guess to conduct optimization at second zoning level,
which includes three zones.
Optimization methodology had been discussed in length already, hence we simply quote the results obtained.
Optimal set of patches is presented on Fig.~\ref{fig::hs-patches-level2}, where
the relative patch widths are correctly preserved.
Corresponding stacking sequences, which are derived from the determined patch structure
via two-step linear integer programming are
\beq
\label{hs-level2}
\begin{array}{cc}
 \mbox{zone 1}: & [ \,5\,/\, 3 \,/\, 3\,]^{(2)}_s \\
    & [0/\mns{45}/0/45/0/\mns{45}/90_3/45/0/\mns{45}/0/45]^{(2)}_s \\
    & \\
 \mbox{zone 2}: & [ \, 6\, /\, 5 \,/\, 4 \,]^{(2)}_s \\
    & [0/45_2/0_2/45/0/\mns{45}_2/90/\mns{45}_2/90_3/45/0/\mns{45}/0/45]^{(2)}_s \\
    & \\
 \mbox{zone 3}: & [ \, 4\,/\, 4 \,/\, 5 \, ]^{(2)}_s \\
    & [0/45/0/\mns{45}/90/45/90/\mns{45}_2/90_3/45/0/\mns{45}/0/45]^{(2)}_s
\end{array}
\eeq
while the corresponding reserve factors and optimal mass are given by
\beq
\begin{array}{c}
M^{(2)} ~=~ 82.5 ~ \mbox{lb}, \\
RF^{(2)} ~=~ (\,1.15, \,\, 1.02,\,\, 1.14 \,)
\end{array}
\eeq

Solution at third level is obtained similarly, corresponding optimal patches are shown on Fig.~\ref{fig::hs-patches-level3},
again with preservation of relative patch widths, stacking sequences for all 9 zones are
\beq
\label{hs-level3}
\begin{array}{cc}
 \mbox{zone 1}: & [\, 3 \,/\, 3 \,/\, 4 \,]^{(3)}_s \\
& [90/\mns{45}/0_3/45/90_2/\mns{45}_2/90/45_2]^{(3)}_s \\[1.5mm]
 \mbox{zone 2}: & [\, 2 \,/\, 3 \,/\, 2 \,]^{(3)}_s \\
& [90/\mns{45}/0/\mns{45}_2/90/45/0/45_2]^{(3)}_s \\[1.5mm]
 \mbox{zone 3}: & [\, 4 \,/\, 4 \,/\, 2 \,]^{(3)}_s \\
& [90/\mns{45}/0/\mns{45}/0/\mns{45}_2/0/45/90/45/0/45_2]^{(3)}_s \\[1.5mm]
 \mbox{zone 4}: & [\, 7 \,/\, 4 \,/\, 2 \,]^{(3)}_s \\
& [90/\mns{45}_2/0/\mns{45}/0_2/\mns{45}/0_3/45/90/45/0/45_2]^{(3)}_s \\[1.5mm]
 \mbox{zone 5}: & [\, 6 \,/\, 5 \,/\, 4 \,]^{(3)}_s \\
& [90/45/90/\mns{45}_2/0/\mns{45}/0/45/90/\mns{45}/0/\mns{45}/0_3/45/90/45_2]^{(3)}_s \\[1.5mm]
 \mbox{zone 6}: & [\, 4 \,/\, 4 \,/\, 4 \,]^{(3)}_s \\
& [90/\mns{45}_2/0/45/90/\mns{45}_2/0_3/45/90_2/45_2]^{(3)}_s \\[1.5mm]
 \mbox{zone 7}: & [\, 5 \,/\, 4 \,/\, 3 \,]^{(3)}_s \\
& [90/\mns{45}/0/\mns{45}_2/0/\mns{45}/0_3/45/90_2/45_3]^{(3)}_s \\[1.5mm]
 \mbox{zone 8}: & [\, 2 \,/\, 3 \,/\, 2 \,]^{(3)}_s \\
& [90/\mns{45}_3/0_2/45/90/45_2]^{(3)}_s \\[1.5mm]
 \mbox{zone 9}: & [\, 2 \,/\, 4 \,/\, 3 \,]^{(3)}_s \\
& [90/45/0/\mns{45}_2/90_2/\mns{45}_2/0/45_3]^{(3)}_s
\end{array}
\eeq
while the corresponding $\rf$ factors and optimal mass are given by
\beq
\label{hs-level3-mass}
\begin{array}{c}
M^{(3)} ~=~ 70.3 ~ \mbox{lb}, \\
RF^{(3)} ~=~ (\, 1.14, \,\, 1.05, \,\, 1.76, \,\, 1.06, \,\, 1.07, \,\, 1.09, \,\, 1.02, \,\, 1.11, \,\, 1.00 \,)
\end{array}
\eeq
As far as the solution at last zoning level is concerned, we found that it is not much different from
(\ref{hs-level3}, \ref{hs-level3-mass}) (mass improvement is less than 2\%), but is significantly more expensive
to obtain. Therefore, in the horseshoe test case reasonable, but yet efficient in mass reduction zoning
structure corresponds to level 3 of our scheme (9 zones), consideration of smaller length scales
seems to be not justified.

A few comments are now in order. First important observation is that optimal mass
(\ref{hs-level3-mass}), obtained in explicit account of all blending rules, is only 13\%
worse than the corresponding lower bound (\ref{hs-mass-th}) and is slightly below
of the result of bi-level approach, Eq.~(\ref{hs-mass-bi}).
This fact becomes even more encouraging if we remind that it corresponds to twice coarser
zone resolution, for which the mass is expected to be larger.
Simultaneously, it is apparent that presented solution still possesses further improvement potential.
Indeed, reserve factors for zones 1 and 8 are $\sim 10\%$ larger than required, while
in zone 3 excess is as large as 75\%. It is true that gap of order 10\% in $\rf$ values
is seen already at level 2 solution and hence could be explained by some deficiencies
of yet unfinished optimization methodology used to solve horseshoe example.
However, even with this reservation zone 3 result deviates too much.
We suspect that this is a direct consequence of blending rules imposition. Indeed, inspection
of Fig.~\ref{fig::hs-general} reveals that loads for zones 2 and 4 significantly differ and
this is reflected in notable difference in corresponding stacking sequences.
However, blending rules require zone 3 to be in-between its neighbors in terms of ply counts
and this is unrelated to actual loads applied to it.
Then it might easily happen that corresponding $\rf$ factor is far beyond imposed bound
and we suggest that this is the reason of observed 75\% excess.

\subsection{Wing Box Model}
\label{section::wb}

\begin{figure}[t]
\centerline{\includegraphics[width=0.75\textwidth]{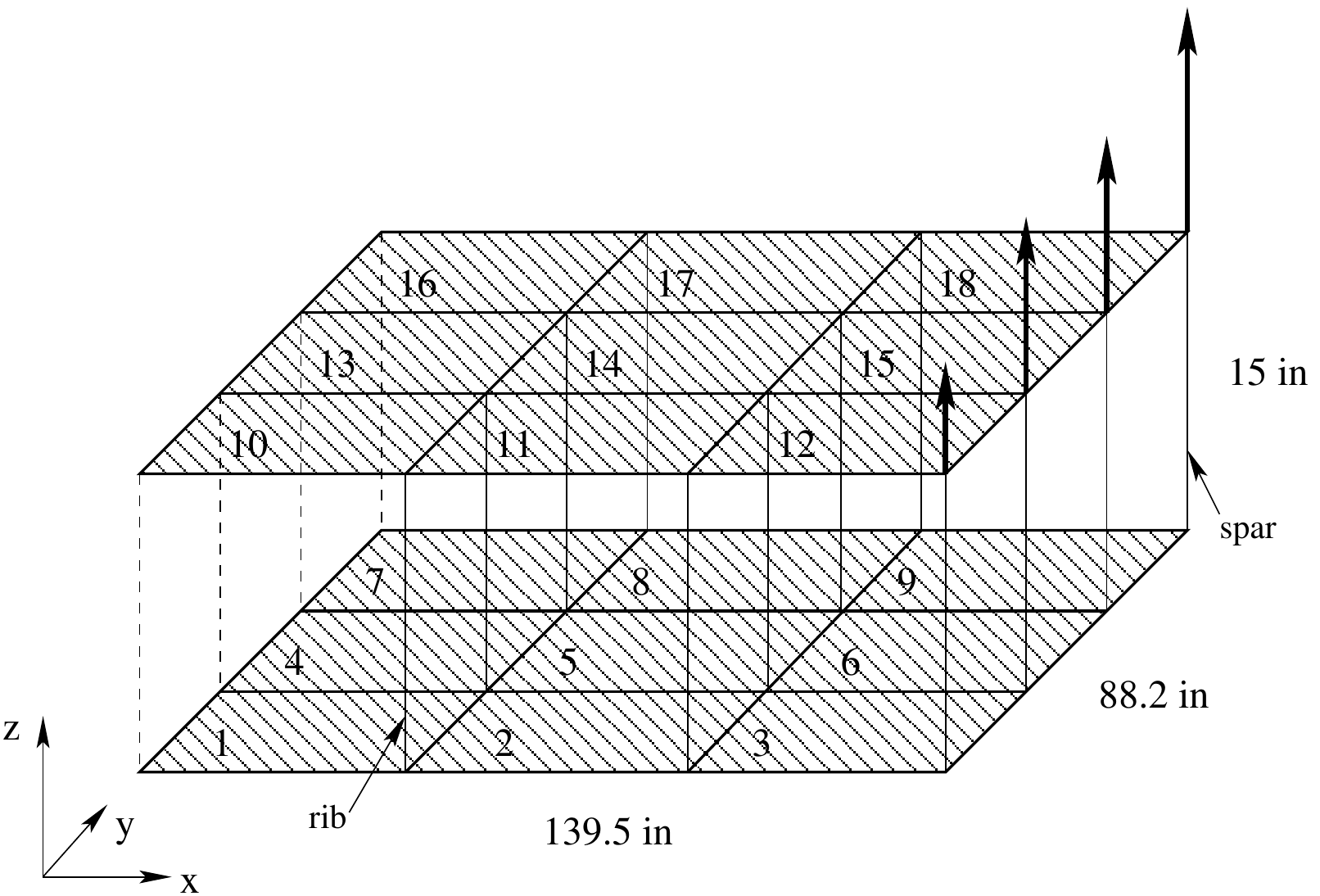}}
\caption{Wing box test case: geometry and setup of the problem.}
\label{fig::wb-general}
\end{figure}

As the second application example we consider well-known wing box test problem, which was
extensively studied in the past~\cite{toropov1,toropov2,Liu-phd,wb1}. 
Problem setup is shown on Fig.~\ref{fig::wb-general}: there are four spars and three ribs
in wing box structure, which are covered from bottom and top sides with 9+9 composite panels,
optimal stacking sequences of which are to be determined. Wing box is clamped at the root,
external loads are applied at the tip as shown on the figure. Material properties, allowed
stains and safety factors are taken the same as in \cite{wb1}, mass of the structure
is obtained using fixed thickness of one ply $0.005~in$ and density $0.057$ lb/in$^3$.
However, compared to Ref.~\cite{wb1} we calculate buckling reserve factors differently
following the analysis of Refs.~\cite{buckling-eigen1, buckling-eigen2} (details are provided
in Appendix \ref{section::appendix}). 
Although it might be guessed in advance that active set of constraints corresponds to strains
for bottom panels and buckling for top, we used generic definition of zone specific reserve
factor (\ref{zone-rf}) with minimum taken with respect to all load types.
Static parts of the structure (ribs and spars) are fixed and have 22 alternating $\pm 45$ plies
with the same material properties.

\begin{figure}[t]
\centerline{\includegraphics[width=0.75\textwidth]{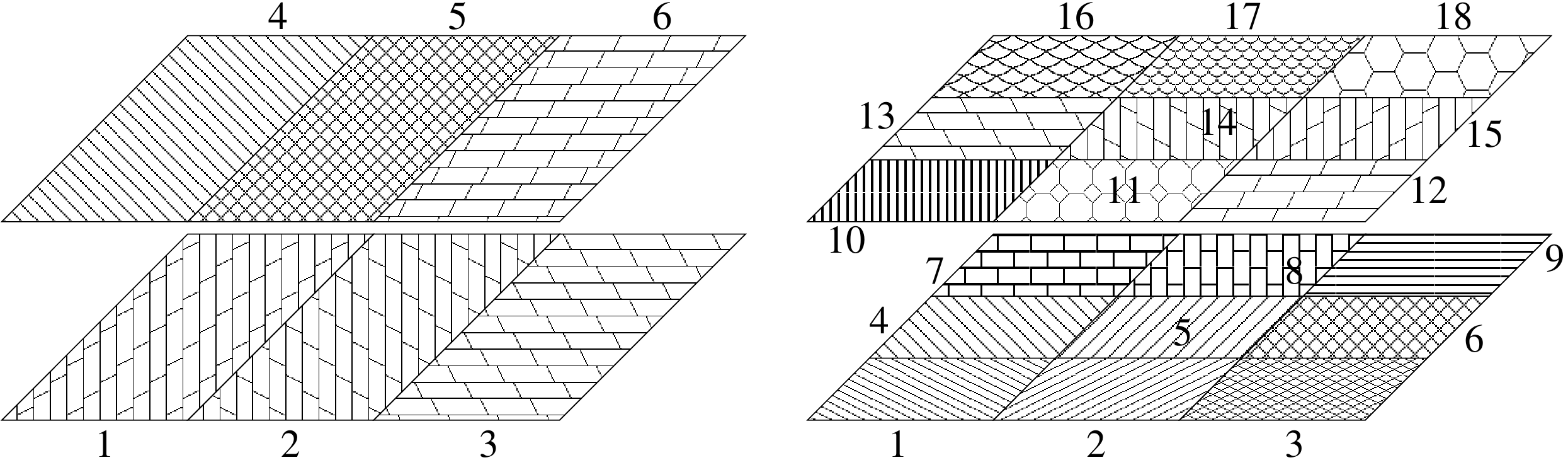}}
\caption{Wing box test case: hierarchical zoning at second (left) and third (right) levels.
First level consists of two zones covering entire bottom and top sides.}
\label{fig::wb-zoning}
\end{figure}

To complete the problem setup it remains to discuss technical aspects of the solution.
It is based on the \pS integration platform, which provides required highly configurable optimization 
algorithms and allows easy integration with ANSYS Mechanical modeling tool used in our study.
As for the later, it had been configured via appropriate APDL scripts to operate in two regimes:
either using preintegrated stiffness matrices given in terms of macroscopic parameters
(upper level of bi-level scheme) or evaluating the model for given complete specification of
relevant stacking sequences.
Mesh size for each zone had been decided during the preliminary problem investigation, we found that
it makes no much sense to consider grids with more than 10 elements along the subzone edges.

Utilized hierarchical zoning structure, Fig.~\ref{fig::wb-zoning},
consists of 3 levels with respective number of zones
2, 6 and 18 (in later case zones coincide with elementary subzones and are indexed the same).
Silent peculiarity of zoning in wing box example is that there are always two disconnected set
of zones at top and bottom sides, which slightly simplifies the treatment.
Selection of particular zoning scheme is dictated by the problem itself and follows from
the qualitative analysis of applied loads.

\begin{table}[t]
\begin{center}
{\small \begin{tabular}{|c|c|c|c|c|c|} \hline
1   & 2   & 3   & 4   & 5   & 6  \\ \hline
$[29/6/4]_s$ & $[17/4/2]_s$ & $[3/3/1]_s$ & $[32/7/4]_s$ & $[19/4/2]_s$ & $[6/3/3]_s$  \\
1.09 & 1.05 & 1.0 & 1.09 & 1.03 & 1.11 \\ \hline \hline
7  & 8   & 9  & 10   & 11 & 12 \\ \hline
$[31/7/4]_s$ & $[18/4/3]_s$ & $[5/4/1]_s$ & $[49/11/9]_s$ & $[29/11/14]_s$ & $[4/13/14]_s$ \\
1.0 & 1.01 & 1.03 & 1.0 & 1.02 & 1.01  \\ \hline \hline
13   & 14  & 15   & 16   & 17   & 18  \\ \hline
$[42/13/15]_s$ & $[21/14/17]_s$ & $[6/14/17]_s$ & $[29/12/29]_s$ & $[18/15/19]_s$ & $[4/14/17]_s$ \\
1.01 & 1.0 & 1.0 & 1.02 & 1.08 & 1.0 \\ \hline
\end{tabular}}
\end{center}
\caption{Wing box test case: zone signatures and $\rf$ factors at third zoning level (no blending rules applied)}
\label{table::wb-level3-th-results}
\end{table}

As in the previous example, we first solved the problem without blending rules imposition
to get the estimate of best attainable mass.
Contrary to the horseshoe problem solution is to be conducted with sequentially refined zoning,
at the first level with only two 
zones the optimal mass and $\rf$ factors are found to be
\beq
\label{wb-level1-th-results}
\begin{array}{rcc}
\mbox{bottom}: & [\,34\,/\,8\,/\,4\,]^{(1)}_s & \rf ~=~ 1.09 \\
\mbox{top}:    & [\,37\,/\,19\,/\,10\,]^{(1)}_s & \rf ~=~ 1.01
\end{array}
\eeq
\beq
\label{wb-level1-th-mass}
M^{(1)}_{th} ~=~ 974.8~\mbox{lb}.
\eeq
Note that even with two zones the problem does not factorize because of additional structural
elements (spars and ribs), which mediate the interactions. Analogous solution at the second zoning level
turns out to be
\beq
\label{wb-level2-th-results}
\mbox{bottom}:\quad
\begin{array}{lll}
[\,33 \,/\, 7 \,/\, 4 \,]^{(2)}_s &  [\, 19 \,/\, 4 \,/\, 2 \,]^{(2)}_s & [\, 6 \,/\, 3 \,/\, 3 \,]^{(2)}_s \\
\rf = 1.01 & \rf = 1.01 & \rf = 1.14
\end{array}
\eeq
\beq
\mbox{top}:\quad
\begin{array}{lll}
[\, 43 \,/\, 15  \,/\, 12 \,]^{(2)}_s &  [\, 20 \,/\, 14  \,/\, 20 \,]^{(2)}_s & [\, 7 \,/\, 14 \,/\, 16 \,]^{(2)}_s \\
\rf = 1.02 & \rf = 1.02 & \rf = 1.05
\end{array}
\eeq
where for each sequentially indexed zone we provide vertically placed ply counts and $\rf$ factors.
Corresponding optimal mass at the second zoning level is
\beq
\label{wb-level2-th-mass}
M^{(2)}_{th} ~=~ 699.0~\mbox{lb}.
\eeq
Finally, at finest possible resolution, in which zones are of minimal allowed extent, established
optimal solution is summarized in Table~\ref{table::wb-level3-th-results}, corresponding total mass is
\beq
\label{wb-level3-th-mass}
M^{(3)}_{th} ~=~ 666.7~\mbox{lb}.
\eeq

A few comments are now in order. One might expect that quality of the solution could be characterized 
by the activity of $\rf$-constraints, it is tempting to conclude that solution is indeed optimal if
it has $\rf \gtrsim 1$  for all zones. Apparently, even with no blending rules applied solution 
presented in (\ref{wb-level3-th-mass}) and Table~\ref{table::wb-level3-th-results}
seems to be slightly suboptimal: in zone 6 (bottom panel) $\rf$ factor exceeds the limiting
value by 10\%. We experimented a lot with this observation and found that judgement of solution quality 
by the activity of $\rf$-type constraints is, in fact, too simplified and should not be taken verbosely.
In particular, in the considered case we found a few other solutions with $\rf_\mu \approx 1$, $\forall\mu$,
but with mass larger by about 5\%. The key issue here is the load redistribution among different zones,
which precludes some $\rf$ factors to become close to 1.
Indeed, in the presented solution diminishing ply counts in zone 6 brings its $\rf$ factor down, but
simultaneously increases the loads on a few other zones, which then are to be made thicker in order
to maintain solution feasibility.

Next we consider wing box problem with the same sequence of zoning levels and explicit reconstruction of
stacking sequences, admissible with respect to all blending rules.
This essentially reduces to solving the problem anew with reconstruction methodology of Section~\ref{section::SLB}
and using different set of APDL scripts. 
Solution at the first zoning level is rather straightforward, it does not include patch construction
algorithm. Corresponding results are
\beq
\begin{array}{c}
\mbox{bottom}: \qquad [\,36\,/\,8\,/\,5\,]^{(1)}_s \qquad \rf ~=~ 1.06 \\[1.5mm]
[45/0/\mns{45}/0_2/\mns{45}/0_3/45/0_3/\mns{45}/0_3/\mns{45}/0_3/\mns{45}/0_3/ \\
\mns{45}/0_3/\mns{45}/0_3/45/0_3/45/0_3/45/90_2/45/0_3/45/0_3/\mns{45}/90_3/45]^{(1)}_s
\end{array}
\eeq
\beq
\begin{array}{c}
\mbox{top}: \qquad [\,45\,/\,16\,/\,7\,]^{(1)}_s \qquad \rf ~=~ 1.08 \\[1.5mm]
[0/45/0_3/\mns{45}/0_3/45/0_3/\mns{45}/0_3/\mns{45}/0_3/45/0_3/\mns{45}/0_3/45/0_3/45/0_3/45/0_2/45/0_3/ \\
\mns{45}/0_3/\mns{45}/0_3/\mns{45}/0_2/45/0/45_3/0/\mns{45}_3/0/45_3/0/\mns{45}_3/90/\mns{45}_3/90_3/45/90_3/45]^{(1)}_s
\end{array}
\eeq
\beq
M^{(1)} ~=~ 988.8~\mbox{lb},
\eeq
which, as expected, are quite similar to (\ref{wb-level1-th-results}).

\begin{figure}[t]
\centerline{\includegraphics[width=0.7\textwidth]{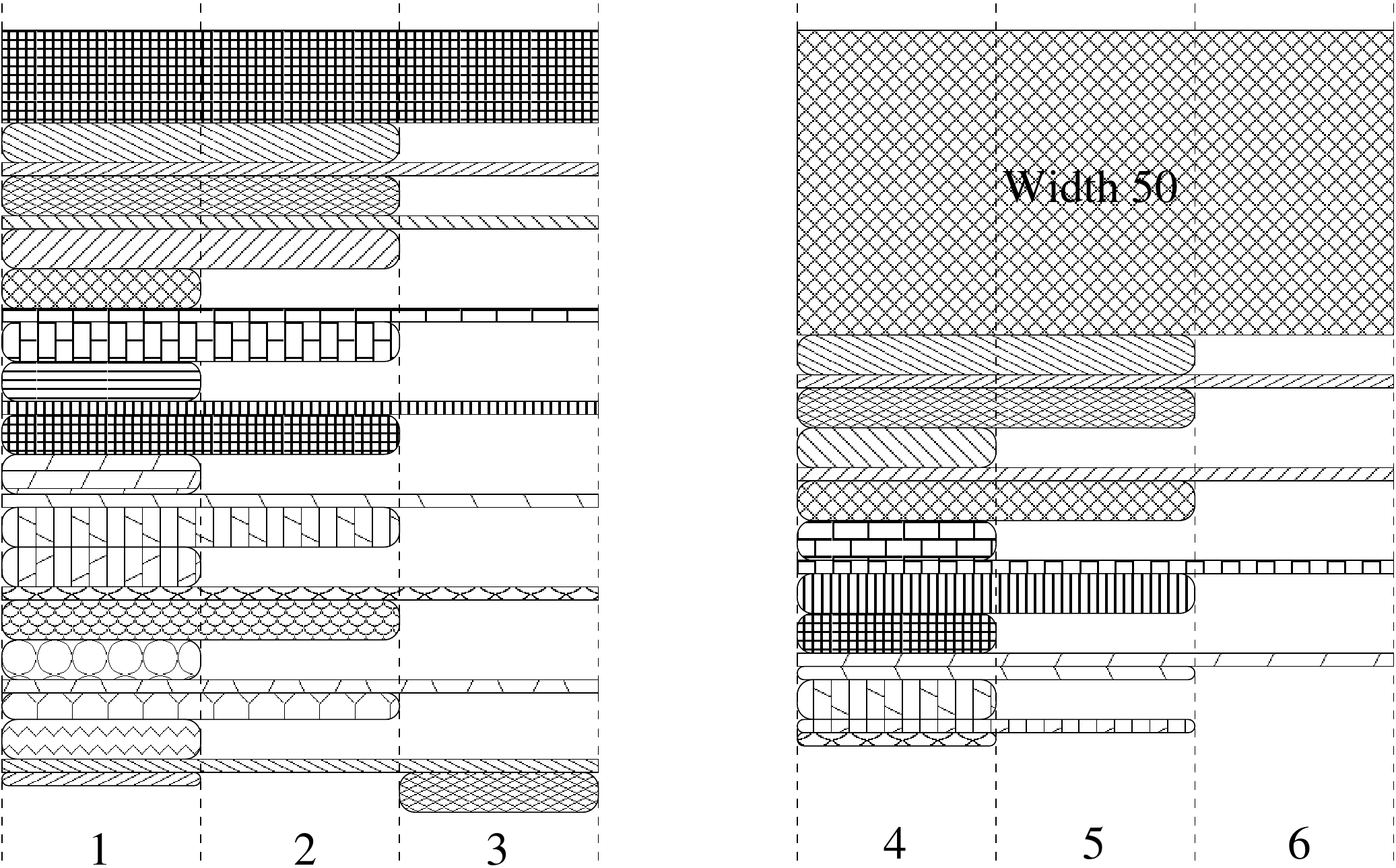}}
\caption{Wing box test case: patch structure of bottom (left) and top (right) sides at second zoning level.}
\label{fig::wb-level2-patches}
\end{figure}

At the next zoning level with $2 \cdot 3$ zones blending rules come into force and dictate the patch
structure presented on Fig.~\ref{fig::wb-level2-patches}, where, as before, relative patch widths are preserved.
Respective solution in terms of zone signatures and resulting $\rf$-factors reads
\beq
\mbox{bottom}:\quad
\begin{array}{lll}
[\,34 \,/\, 9 \,/\, 5 \,]^{(2)}_s &  [\, 19 \,/\, 7 \,/\, 5 \,]^{(2)}_s & [\, 5 \,/\, 4 \,/\, 5 \,]^{(2)}_s \\
\rf = 1.01 & \rf = 1.08 & \rf = 1.15
\end{array}
\eeq
\beq
\label{wb-level2-results}
\mbox{top}:\quad
\begin{array}{lll}
[\, 29 \,/\, 20  \,/\, 12 \,]^{(2)}_s &  [\, 21 \,/\, 18  \,/\, 11 \,]^{(2)}_s & [\, 13 \,/\, 15 \,/\, 11 \,]^{(2)}_s \\
\rf = 1.0 & \rf = 1.03 & \rf = 1.02
\end{array}
\eeq
with corresponding mass value being
\beq
\label{wb-level2-mass}
M^{(2)} ~=~ 738.7~\mbox{lb},
\eeq
while per zone stacking sequences are summarized in Table~\ref{table::wb-level2-ss}.
Compared to (\ref{wb-level2-th-results}, \ref{wb-level2-th-mass}) one notice that solution remains qualitatively
the same, both in terms of total ply count and $\rf$ factors.
Mass difference is mostly because of thicker bottom panels, while top side is characterized by redistribution
of plies from root to tip, which is a consequence of applied blending rules.

\begin{table}[t]
\begin{center}
{\small \begin{tabular}{|c|c|c|c|c|c|} \hline
1   & 2   & 3   & 4   & 5   & 6  \\ \hline
$[27/9/5]_s$ & $[19/6/3]_s$ & $[6/3/2]_s$ & $[30/9/5]_s$ & $[16/6/3]_s$ & $[6/3/2]_s$ \\
1.01 & 10.7 & 1.26 & 1.06 & 1.04 & 1.25 \\ \hline \hline
7  & 8   & 9  & 10   & 11 & 12 \\ \hline
$[36/12/6]_s$ & $[20/6/5]_s$ & $[6/5/7]_s$ & $[30/16/19]_s$ & $[13/19/12]_s$ & $[14/12/8]_s$ \\
1.03 & 1.02 & 1.3 & 1.14 & 1.0 & 1.21 \\ \hline  \hline
13   & 14  & 15   & 16   & 17   & 18  \\ \hline
$[32/16/19]_s$ & $[14/18/19]_s$ & $[18/13/13]_s$ & $[26/17/24]_s$ & $[15/20/11]_s$ & $[22/11/7]_s$ \\
1.0 & 1.01 & 1.22 & 1.02 & 1.02 & 1.0 \\ \hline
\end{tabular}}
\end{center}
\caption{Wing box test case: zone signatures and $\rf$ factors at third zoning level (optimal values).}
\label{table::wb-level3-results}
\end{table}

Solution at the third level is constructed similar to above, however, corresponding patch structure is
difficult to present because of two-dimensional connectivity structure of zones and large number
of patches. It suffices to note that there are 36 patches at bottom side, 8 of which are continuous with
outer-most layer having final width 4 (all continuous patches placed internally are of unit width).
At top side solution has 86 patches, 23 of which are continuous and outer-most layer is of width 11 (similar to
bottom case all other continuous patches have unit width).
Signatures of optimal stacking sequences for level 3 solution, as well as respective
reserve factors are summarizes in Table~\ref{table::wb-level3-results}, while
Table~\ref{table::wb-level3-stack} presents complete list of stacking sequences for all zones.
Optimal mass is found to be
\beq
\label{wb-level3-mass}
M^{(3)} ~=~ 718.5~\mbox{lb}.
\eeq
Comparison with corresponding level 3 results in case of no blending rules imposition reveals that
qualitative differences, noted already for level 2, remain the same.
Bottom plate becomes thicker, while top side of wing box remains similar. 
Apparent feature of the presented solution is that almost all zones, closest to tip, have sufficiently
large $\rf$ values. This is dictated by imposed blending rules: without them the width of both bottom and top
sides rapidly diminishes from root to tip of the structure, however, large width changes are effectively
forbidden by the dropping rule (among other factors).

Perhaps, the most important is to note that the difference between optimal masses
(\ref{wb-level3-th-mass}) and (\ref{wb-level3-mass}) is about 8\%.
This means, in particular, that
blending rules imposition do not lead to severe non-localities in zone interactions.
Indeed, one might expect that continuity and dropping rules could effectively induce additional
non-local interactions  between different zones.
However, effect seems to be parametrically small and does not result in drastic efficiency degradation.

\begin{table}
\begin{center}
{\small\begin{tabular}{|r|l|} \hline
1\rule[-4pt]{0pt}{18pt}
  & $[0/\mns{45}/0_2/\mns{45}/0_3/45/0_2/\mns{45}/0_3/45/0_2/\mns{45}/90_2/\mns{45}/0_3/$ \\
  & $\mns{45}/0_3/\mns{45}/0_2/45/0_2/45/0_3/45/0_3/45/0_2/45/0_2/45/90_2/\mns{45}/90/\mns{45}/0/45]_s$ \\
2\rule[-4pt]{0pt}{18pt}
  & $[\mns{45}/0_3/\mns{45}/0_3/\mns{45}/90_2/\mns{45}_2/0_3/45/0_2/45_2/0_3/45/0_2/45/0_2/45/90_2/\mns{45}/90/\mns{45}/0/45]_s$ \\
3\rule[-4pt]{0pt}{18pt}
  & $[45/90_2/\mns{45}/0_2/\mns{45}/0/45/0/45/90_2/\mns{45}/90/\mns{45}/0/45]_s$ \\
4\rule[-4pt]{0pt}{18pt}
  & $[90/45_3/0_2/45/0_3/45_2/0_2/\mns{45}/0_2/\mns{45}/0_3/\mns{45}/0_2/\mns{45}/0_2/\mns{45}/0_2/\mns{45}/90_2/45/0_3/\mns{45}/0_2/\mns{45}/$ \\
  & $90_3/\mns{45}/0_3/\mns{45}_2/90/\mns{45}/90/45_3/0/45_3/0/\mns{45}_3/90/45_2/90/\mns{45}_2/90/\mns{45}_3/0/45_3/90/45_2]_s$ \\
5\rule[-4pt]{0pt}{18pt}
  & $[45/0/45_3/0_2/\mns{45}/0_3/\mns{45}/0_2/\mns{45}/0_2/\mns{45}/90_2/45/0_3/\mns{45}/0_2/\mns{45}/90_3/\mns{45}/0_3/\mns{45}_2/$ \\
  & $90/\mns{45}/90/45_3/0/45_3/0/\mns{45}_3/90/45_2/90/\mns{45}_2/90/\mns{45}_3/0/45_3/90/45_2]_s$ \\
6\rule[-4pt]{0pt}{18pt}
  & $[45/0_2/\mns{45}/90_2/45/0_3/\mns{45}/0_2/\mns{45}/90_3/\mns{45}/0_3/\mns{45}_2/90/\mns{45}/90/45_3/0/45_3/$ \\
  & $0/\mns{45}_3/90/45_2/90/\mns{45}_2/90/\mns{45}_3/0/45_3/90/45_2]_s$ \\ \hline
\end{tabular}}
\end{center}
\caption{Wing box test case: optimal stacking sequences at second zoning level.}
\label{table::wb-level2-ss}
\end{table}

\begin{table}
\begin{center}
{\small \begin{tabular}{|r|l|} \hline
1\rule[-4pt]{0pt}{18pt}
   & $[45/0/45/0/45/0_3/45/90/\mns{45}/0/\mns{45}/90_2/\mns{45}/0_2/\mns{45}/0_2/\mns{45}/0_3/\mns{45}/0_3/45/90_2/\mns{45}/$ \\
   & $0_2/\mns{45}/0/45/0_2/45/0_3/45/0/\mns{45}/0_2/45]_s$ \\
2\rule[-4pt]{0pt}{18pt}
   & $[0/45/0/45/0_2/45/90_3/\mns{45}/0_2/\mns{45}_2/0_3/\mns{45}/0_2/\mns{45}/0/45/0_3/45/0_2/\mns{45}/0_2/45]_s$ \\
3\rule[-4pt]{0pt}{18pt}
   & $[45/90_2/\mns{45}/0/\mns{45}/0_2/45/0/\mns{45}/0_2/45]_s$ \\
4\rule[-4pt]{0pt}{18pt}
   & $[0/45/0/45/0/45/0_3/45/90/\mns{45}/0/\mns{45}/90_2/\mns{45}/0_2/\mns{45}/0_2/\mns{45}/0_3/\mns{45}/0_3/45/90_2/\mns{45}/$ \\
   &  $0_2/\mns{45}/0/45/0_2/45/0_3/45/0_3/\mns{45}/0_2/45]_s$ \\
5\rule[-4pt]{0pt}{18pt}
   & $[0/45/0/45/0_2/45/90_2/\mns{45}/0_2/\mns{45}_2/0/\mns{45}/0_2/\mns{45}/0/45/0_3/45/0/\mns{45}/0_2/45]_s$ \\
6\rule[-4pt]{0pt}{18pt}
   & $[45/90_2/\mns{45}/0/\mns{45}/0_2/45/0/\mns{45}/0_2/45]_s$ \\
7\rule[-4pt]{0pt}{18pt}
   & $[90/45_2/0/45/0/45/0_2/45/90/\mns{45}/0/\mns{45}/90_2/\mns{45}/0_2/\mns{45}/0_2/\mns{45}/0_2/\mns{45}/0_3/\mns{45}/0_3/45/$ \\
   & $0_2/45/90_2/\mns{45}/0_2/\mns{45}/0_2/\mns{45}/0/45/0_2/45/0_2/45/0_3/45/0_3/\mns{45}_2/0_2/45]_s$ \\
8\rule[-4pt]{0pt}{18pt}
   & $[0/45/90_3/\mns{45}/0_2/\mns{45}/0_3/\mns{45}/0_3/45/90_2/\mns{45}/0_2/\mns{45}/0/45/0_2/45/0_3/45/0/\mns{45}/0_2/45]_s$ \\
9\rule[-4pt]{0pt}{18pt}
   & $[90/\mns{45}/0/45/90_2/45/90_2/\mns{45}_2/0/45/90_2/\mns{45}/0/45/0/\mns{45}/0_2/45]_s$ \\
10\rule[-4pt]{0pt}{18pt}
   & $[\mns{45}/90_3/\mns{45}/90_2/45/90_2/45/90_2/\mns{45}/0_2/45/0_2/\mns{45}/90_2/45/0_2/45/0_2/\mns{45}/0_3/45/0/$ \\
   & $45/0_3/45/0_2/\mns{45}_2/0_3/45/90_2/45/90/45/0_3/45/0/\mns{45}_2/0/\mns{45}/0_3/\mns{45}/90_2/\mns{45}/90_2/$ \\
   & $\mns{45}/0/45/90/\mns{45}_3/0/45_3]_s$ \\
11\rule[-4pt]{0pt}{18pt}
   & $[\mns{45}/90/\mns{45}/90_2/45/90/45_2/0/\mns{45}_2/0_2/\mns{45}/90_2/45_2/0/45_3/0/\mns{45}/0/45_2/0/\mns{45}/90/$ \\
   & $\mns{45}_3/90/\mns{45}/0/45/90/45_2/90/45_2/0/\mns{45}_2/0/\mns{45}/0/\mns{45}/90/\mns{45}/0/45/90/\mns{45}_3/0/45_3]_s$ \\
12\rule[-4pt]{0pt}{18pt}
   & $[\mns{45}/90_3/45_2/0_2/45/0_2/\mns{45}/0/45_2/0/\mns{45}_2/0/45_3/0/\mns{45}/0/\mns{45}/0_3/\mns{45}/90_2/\mns{45}/$ \\
   & $90_2/\mns{45}/0/45/90/\mns{45}_3/0/45_3]_s$ \\
13\rule[-4pt]{0pt}{18pt}
   & $[0/\mns{45}_2/90/\mns{45}/90/\mns{45}/90_3/45/90_2/45/90_2/\mns{45}/0/45/0_3/45/0_3/45/0_2/\mns{45}/0_3/45/0/45/$ \\
   & $0_3/45/0_2/\mns{45}/0_3/45/90_2/45/90/45/0_3/45/0_2/\mns{45}/90_2/\mns{45}/0_3/\mns{45}/90_2/\mns{45}_2/90_2/$ \\
   & $\mns{45}/0/45/90/\mns{45}_3/0/45_3]_s$ \\
14\rule[-4pt]{0pt}{18pt}
   & $[\mns{45}/90_2/\mns{45}/90_2/\mns{45}_2/90_2/45/90_3/45_2/0_2/\mns{45}/90_2/45/0/45_3/0/\mns{45}/0/45_2/0/\mns{45}_2/90/$ \\
   & $\mns{45}/0/45/90/45_2/90/45_2/0/\mns{45}_2/0/\mns{45}/0_3/\mns{45}/90_2/\mns{45}/90_2/\mns{45}/0/45/90/\mns{45}_3/0/45_3]_s$ \\
15\rule[-4pt]{0pt}{18pt}
   & $[\mns{45}/90/\mns{45}_2/90_2/45/90_2/45/0_2/45/0_2/\mns{45}/0/45_2/0/\mns{45}/0/45/90/45/0_3/45/0/\mns{45}/$ \\
   & $90_2/45/0_3/\mns{45}/90_2/\mns{45}/0_2/\mns{45}/90_2/\mns{45}/0/45/90/\mns{45}_3/0/45_3]_s$ \\
16\rule[-4pt]{0pt}{18pt}
  & $[45/0_2/45_2/90/\mns{45}/90_3/\mns{45}/0/45/90_2/\mns{45}/90_2/\mns{45}/0/\mns{45}/90_3/45/90_2/45/0_2/45/0_3/$ \\
  & $\mns{45}/0_2/45_2/0_2/\mns{45}/0_3/45/90_2/45/90/45/0_3/45/0/\mns{45}_2/0/\mns{45}/0_3/\mns{45}/90_2/\mns{45}/$ \\
  & $90_3/\mns{45}/90_2/\mns{45}/0/45/90/\mns{45}_3/0/45_3]_s$ \\
17\rule[-4pt]{0pt}{18pt}
  & $[0/\mns{45}_2/90/\mns{45}/90_2/45_3/90/\mns{45}/0/\mns{45}/90/\mns{45}/0/45/0/45_3/0_2/\mns{45}_2/0/45/0/45_3/0/$ \\
  & $\mns{45}_2/90/\mns{45}/0/45/90/45_2/90/45_2/0/\mns{45}_2/0/\mns{45}/0/\mns{45}_2/90_2/\mns{45}/0/45/90/\mns{45}_3/0/45_3]_s$ \\
18\rule[-4pt]{0pt}{18pt}
  & $[\mns{45}/90_3/45/0_2/45/0_3/45/0/45_2/0_3/\mns{45}/0_2/\mns{45}_2/0/45/0_3/45/0_2/\mns{45}/0_3/\mns{45}/$ \\
  & $90_2/\mns{45}/90/\mns{45}/0/45/90/\mns{45}_3/0/45_3]_s$ \\ \hline
\end{tabular}}
\end{center}
\caption{Wing box test case: solutions at third zoning level (optimal stacking sequences)}
\label{table::wb-level3-stack}
\end{table}

\section{Conclusions}
\label{section::consclusion}

In this paper we addressed long standing problem of automated optimization of laminated composite structures
and proposed methodology, which seems to solve major deficiencies of known approaches.
Here we mean primarily the bi-level scheme, which provides clear scales separation, explicitly utilizes
small parameter in the problem and have therefore a great potential efficiency.
However, its application scope remains rather limited.
Ultimate reason of this is that distinguishing macro- and micro-scopic variables, which is the cornerstone
of bi-level method, contradicts widely accepted industrial practices, where modeling tools evaluate
performances and constraints directly from the exhaustive set of stacking sequences.
In a nutshell, one is faced therefore with the problem of computationally inexpensive determination
of relevant stacking sequences, identically fulfilling blending and composition rules,
for given set of macroscopic parameters.

We solved this issue and proposed specific ``greedy'' algorithm of stacking sequences determination,
which is close in spirit, but is vastly different in details, to Shared Layers Blending approach.
In particular, our method performs outer blending construction with built in composition rules.
Algorithm first determines an exhaustive set of patches, for which continuity and dropping rules are
satisfied identically for any choice of ply orientations, and then solves integer linear programming problem,
which warrants continuity and disorientation requirements.

We also provided the construction of tight convex approximation to feasible domain of macroscopic parameters,
which allows to avoid the use of generically less efficient collaborative optimization schemes.
Combined with hierarchical zoning it permitted us to develop optimization methodology, which inherits
prime advantages of bi-level scheme, but is free from its prime deficiencies.

Performance of proposed methodology was tested in two example problems, particular advantage of which
is the possibility to solve them without blending rules imposition. It provides a lower bound on attainable
performance in each case and gives the reference scale to compare our results with.
Comparison reveals that strict accounting for blending rules worsens the results only slightly,
in both cases obtained optimal solutions are within 10\% gap from the respective theoretical lower bounds.

\section*{Acknowledgments}
\label{section::acknowledgments}

The authors are grateful to the members of Applications Engineering Department of DATADVANCE LLC
for stimulating environment.
The thorough discussions with A.Pospelov, A.Saratov are kindly acknowledged.
This work was partially supported by the grant RFBR-15-29-07043.

\appendix
\section{Calculation of Buckling Reserve Factor}
\label{section::appendix}
Here we briefly summarize how critical buckling factor is calculated for simply supported
rectangular plates under the bending and compression loads. Derivation closely follows Ref.~\cite{buckling-eigen2},
therefore we only provide essentially distinct points.

The strain energy of buckled rectangular $a_1 \times a_2$ plate is given by
\beq
\label{strain-energy}
\Delta U ~=~ \frac{1}{2} \int\limits_{a_1 \times a_2} \,\, d^2x \,\, D^{\alpha\beta} \,\,
\partial^2_\alpha w \,\, \partial^2_\beta w\,,\qquad
D^{\alpha\beta} ~=~ \left[ \begin{array}{cc} D^{11} & D^{12} + 2 D^{66} \\ D^{12} + 2 D^{66} & D^{22} \end{array} \right]\,,
\eeq
where $w$ is the orthogonal to plate deflection coordinate, $D^{\alpha\beta}$ is flexural rigidity tensor
and repeated indices taking values $1,2$ are summed up. The work performed by external forces reads
\beq
\label{force-energy}
\Delta T ~=~ -\frac{1}{2} \int\limits_{a_1 \times a_2} \,\, d^2x \,\, N^{\alpha\beta} \,\,
\partial_\alpha w \,\, \partial_\beta w\,,\qquad
N^{\alpha\beta} ~=~ \left[ \begin{array}{cc} N^x & N^{xy} \\ N^{xy} & N^y \end{array} \right]\,,
\eeq
where $N^{\alpha\beta}$ are the in-plane forces per unit length.
Loads criticality is characterized by the appearance of non-trivial solution to $\Delta U + \Delta T = 0$.
Using proper eigenmodes representation for simply supported plates
\beq
w ~=~ \sum_{n_1, n_2} \, w_{n_1, n_2} \, \prod_\alpha \sin[\frac{\pi n_\alpha x_\alpha}{a_\alpha}]
\eeq
one obtains the following equation for the amplitudes $w_{n_1, n_2}$:
\beq
\label{critical-eigen1}
(\pi^2 D^{\alpha\beta} n^2_\alpha n^2_\beta ~+~ N^\alpha n^2_\alpha) w_{n_1, n_2}
~=~ \frac{32}{\pi^2} N^{12} \sum_{m_1, m_2} \frac{1}{2} \left(
\left[\begin{array}{cc} n_1 & n_2 \\ m_1 & m_2 \end{array}\right] +
\left[\begin{array}{cc} m_1 & m_2 \\ n_1 & n_2 \end{array}\right] \right) \,\, w_{m_1, m_2}\,,
\eeq
where
\beq
\left[\begin{array}{cc} p & q \\ n & m \end{array}\right] ~=~ \frac{pqnm}{(p^2 - n^2)\,(q^2 - m^2)} \chi_{p,n} \chi_{q,m}\,,
\qquad
\chi_{p,q} = \left\{ \begin{array}{cl} 0, & p+q = 2k \\ 1, & p+q = 2k+1 \end{array}\,.
\right.
\eeq
Critical loads $N$ correspond to the first encountered zero eigenvalue of the matrix defining the linear
system (\ref{critical-eigen1}). Alternatively, at fixed loads we may consider scaling parameter $\lambda$
(commonly known as buckling reserve factor), at which the loads $\lambda N$ become critical.
Therefore the problem is to determine minimal positive $\lambda$ when lowest eigenvalue of symmetric matrix
\beq
\label{critical-matrix}
Q_{\mu\nu} ~=~ \frac{1}{\lambda} \cdot \pi^2 D^{\alpha\beta} n^2_\alpha n^2_\beta \cdot \delta_{\mu\nu}  ~+~
               N^\alpha n^2_\alpha \cdot \delta_{\mu\nu}  ~-~
               \frac{32}{\pi^2} N^{12} \cdot\frac{1}{2} \left(
                  \left[\begin{array}{c} \nu \\ \mu \end{array}\right] +
                  \left[\begin{array}{c} \mu \\ \nu \end{array}\right]\right )
\eeq
becomes zero (here Greek symbols denote multi-indices, $\mu = (n_1, n_2)$).
Note that for sufficiently small $\lambda$ the matrix $Q$ is surely positive definite, while the sign
of its minimal eigenvalue $\lambda_{min}$ for $\lambda \to +\infty$ depends upon the distribution of external loads.

We calculate buckling reserve factor $\lambda$ via the solution of non-linear equation
$\lambda_{min}( \lambda, N ) = 0$ along the lines presented in Ref.~\cite{buckling-eigen2}.
Namely, the sequence of finite order approximations to the infinite sum in (\ref{critical-eigen1})
(and hence the size of matrix (\ref{critical-matrix})) is considered until respective values of reserve
factor $\lambda$ stabilize.
This provides robust and rather precise estimate of buckling reserve factor applicable in case of general loads.


\end{document}